\documentclass[iop,apj]{emulateapj}

\usepackage{threeparttable}

\graphicspath{{figures/}}
\newcommand{\ud}{\mathrm{d}}

\newcommand{\simgt}{\lower.5ex\hbox{$\; \buildrel > \over \sim \;$}}
\newcommand{\simlt}{\lower.5ex\hbox{$\; \buildrel < \over \sim \;$}}

\shorttitle{Redshift-depth magnification in SDSS/BOSS}
\shortauthors{Coupon et al.}

\begin{document}

\title{Cluster Lensing Profiles Derived from a Redshift Enhancement\\ 
of Magnified BOSS-Survey Galaxies}

\author{Jean Coupon}
\affil{Institute of Astronomy and Astrophysics, Academia Sinica, \\ 
P.O. Box 23-141, Taipei 10617, Taiwan}
\email{coupon@asiaa.sinica.edu.tw}

 \author{Tom Broadhurst\altaffilmark{1}}
 \affil{Department of Theoretical Physics, University of Basque
   Country\\ UPV/EHU, P.O. Box 644, E-48080 Bilbao, Spain}
 \affil{IKERBASQUE, Basque Foundation for Science, Alameda Urquijo 36-5, E-48008 Bilbao, Spain}
 \and

 \author{Keiichi Umetsu}
 \affil{Institute of Astronomy and Astrophysics, Academia Sinica, \\
   P.O. Box 23-141, Taipei 10617, Taiwan}

 \altaffiltext{1}{Institute of Astronomy and Astrophysics, Academia Sinica, P.O.
   Box 23-141, Taipei 10617, Taiwan}

\begin{abstract}
  We report the first detection of a redshift-depth enhancement of
  background galaxies magnified by foreground clusters.  Using
  $300\,000$ BOSS-Survey  galaxies with accurate spectroscopic redshifts, we
  measure their mean redshift depth behind four large samples of optically selected 
  clusters from the SDSS surveys, totalling $5\,000-15\,000$ clusters. A
  clear trend of increasing mean redshift towards the cluster centers is found, averaged over each 
  of the four cluster samples. In addition we find similar but noisier 
  behaviour for an independent X-ray
  sample of 158 clusters lying in the foreground of the current BOSS
  sky area.  By adopting the mass-richness relationships appropriate
  for each survey we compare our results with theoretical predictions
  for each of the four SDSS cluster catalogs.  The radial form of this
  redshift enhancement is well fitted by a richness-to-mass weighted
  composite Navarro-Frenk-White profile with an effective mass ranging between
  $M_{200}\sim1.4$--$1.8\times10^{14}M_\sun$ for the optically
  detected cluster samples, and $M_{200}\sim5.0\times10^{14}M_\sun$
  for the X-ray sample.  This lensing detection helps to establish the credibility of these
  SDSS cluster surveys, and provides a normalization for their respective 
  mass-richness relations. In the context of the
  upcoming bigBOSS, Subaru-PFS, and EUCLID-NISP spectroscopic surveys, this method
  represents an independent means of deriving the masses of
  cluster samples for examining the cosmological evolution, and provides 
  a relatively clean consistency check of weak-lensing measurements,
  free from the systematic limitations of shear calibration.

\end{abstract}


\section{Introduction}
   
Our current paradigm predicts that the
properties of dark energy will be imprinted in the matter power
spectrum and its time evolution. Although probing the large scale
structure in the universe is a promising strategy to investigate the
different dark-energy model candidates, it is becoming increasingly
challenging to detect the relatively subtle signatures of the various
lesser ingredients and higher order effects, forcing us to observe
many independent physical phenomena over gigantic volumes, to
reach sufficient joint statistical precision.

In the era of large scale cosmological surveys, gravitational lensing
is one of the most promising methods for establishing the fluctuations
of matter in the universe, as lensing is purely gravitational and
can be largely assumed to be independent of the unknown nature of the 
dark matter. One may employ  two-dimensional 
\citep{2013MNRAS.tmp..735K} or tomographic
\citep{2012arXiv1212.3327B} \emph{cosmic shear} to measure the
integrated lensing effect of matter fluctuations along the
line-of-sight up to very large scales \citep{2004MNRAS.353.1176T,2007ApJS..172..239M}.
Or, one may use cluster abundances \citep{2010ApJ...708..645R} 
to put constraints on the evolution of the dark
matter halo mass function, which is known to be a very sensitive
function of the evolution of the mean mass density and of the dark
energy properties. Masses can be derived from the radial shear profile around
clusters in the weak lensing regime
\citep{2009ApJ...703.2217S,2010ApJ...709...97L,2011MNRAS.414.1851O,2012ApJ...755...56U}, from X-ray \citep{2008ApJ...675.1106R},
SZ effect \citep{2011ApJ...737...61M}, or satellite kinematics
\citep{2011MNRAS.410..210M}.

Even though lensing from stacked clusters undoubtedly leads to a
less noisy mass estimate among all these methods, accurate statistics
will be limited, however, by our ability to measure galaxy shapes down
to the required precision. Recently,
\citet{2012MNRAS.427..146H} demonstrated that for a survey like the
CFHTLS-wide, spreading over 154~deg$^2$, the level of systematics in
shape measurement, as given by the state-of-the art shape measurement
technique \citep[Lensfit,][]{2013MNRAS.429.2858M},
was lower than statistical errors. For upcoming experiments such as the Hyper Suprime
Cam \citep[HSC,][]{2012SPIE.8446E..0ZM} survey, the Dark Energy Survey
(DES\footnote{\url{http://www.darkenergysurvey.org}}), the 
Large Synoptic Survey Telescope project (LSST\footnote{\url{http://www.lsst.org/}})
or EUCLID\footnote{\url{http://www.euclid-ec.org/}}
 survey \citep{2011arXiv1110.3193L} 
which will probe volumes orders of magnitude larger than
the CFHTLS-wide, the limiting requirements will be much more stringent.

Lens magnification provides independent observational
alternatives and complementary means to weak shape measurement
\citep{2010ApJ...723L..13V,2011ApJ...733L..30H}. In practice
magnification bias has been used to improve the mass profiles of
individual massive clusters, and stacked samples of clusters from
recent dedicated cluster surveys. Here the number counts of flux
limited samples of background sources is reduced in surface density by 
the magnified sky area, and compensated to some extent by fainter objects 
magnified above the flux limit \citep[][hereafter BTP95]{1995ApJ...438...49B}
 with a net magnification-bias in the surface density
depending on the slope of the background counts. For very red
background galaxies behind clusters the net effect is a clear
``depletion" of red background galaxies towards the cluster
center \citep{2005ApJ...619L.143B,2008ApJ...685L...9B}
and is used in combination with weak shear to enhance the precision of individual
cluster mass profiles \citep{2008ApJ...684..177U,2011ApJ...729..127U}.  
A strength of this method resides in the fact that no 
shape information is used and is therefore less prone to 
systematic errors, as opposed to shear
measurement for which complex corrections are required 
\citep{1995ApJ...449..460K,2008ApJ...684..177U}

In the statistical regime, for deep wide area surveys, progress has been
made recently in a similar way using distant Lyman-break galaxies
behind stacked samples of foreground clusters, resulting in a claimed
positive magnification-bias enhancement of the observable surface
density of background galaxies at small radius. In practice, some
cross contamination of cluster members and foreground galaxies with
the background will complicate such measurements. Until recently,
magnification bias studies
\citep{2012ApJ...754..143F,2013MNRAS.429.3230H} have faced such a high
statistical error that their conclusions were not affected by these
systematics.  However in the future a careful treatment of potential
sources of systematics will be mandatory.

With redshift information for large samples of background galaxies,
cluster magnification may be tackled more fully, as the
luminosity function of background galaxies is magnified both in the
density and luminosity directions, in a characteristic redshift
dependent way. For flux limited redshift samples, the redshift
distribution becomes skewed to higher mean redshift as a consequence of
magnification, and the mean luminosity is also modified owing to the 
curvature (non-power law) of the luminosity
function (BTP95). The latter effect has been claimed in the
case of magnified QSO's behind galaxies in the SDSS survey
\citep{2010MNRAS.405.1025M}.

In this paper, we are interested in the enhanced redshift depth of
background galaxies magnified by foreground clusters and averaged over
large cluster samples recently identified within the SDSS
survey. We measure the mean redshift of BOSS galaxies behind SDSS
clusters
\citep{2000AJ....120.1579Y,2009astro2010S.314S,2013AJ....145...10D}, 
using four different catalogues of optically detected clusters found
from very different independent methods, including the maxBCG
\citep{2007ApJ...660..239K}, the GMBCG \citep{2010ApJS..191..254H},
the \citet{2011ApJ...736...21S} and the \citet{2012ApJS..199...34W} samples.
We compare the depth magnification to the expected lensing signal
using mass-richness relationships calibrated from X-ray and weak 
shear measurements. In addition we use the MCXC X-ray cluster sample
assembled by \citet{2011A&A...534A.109P} from a number of 
independent ROSAT observations. 

In Section \S2 we present the data sets used in this study; the
BOSS sample, the four galaxy cluster catalogs and the X-ray sample.
In a third section,
we describe our model, and in a fourth section we present our
measurement technique, followed by our results presented and discussed
in Section \S5. In Section \S6 we present a brief error analysis
and we conclude in Section \S7.  Everywhere we assume a flat
$\Lambda$CDM cosmology with $\Omega_m = 0.258$, $\Omega_{\lambda} =
0.742$, and $h_{100} = 0.72$ \citep{2009ApJS..180..225H}. All masses
are expressed in unit of $M_{\odot}$.

\section{Data}

\subsection{The lens samples}
\label{sec:data-lens}

The  Sloan Digital Sky Survey \citep[SDSS,][]{2000AJ....120.1579Y}
is an optical ($ugriz$ filters) photometric and spectroscopic 
survey which represents the largest probe of the local universe to date.
In addition to its unprecedented volume observed, the success 
of the SDSS project is based upon an accurate photometric calibration, 
an ingenious synergy between photometric and spectroscopic 
observing strategies, and a handy and well documented database.
In the published literature, several methods have been used to 
construct cluster samples from the SDSS database. 
In this study, we thus focus on the following four 
publicly-available cluster catalogs based on optical identifications.

First, the maxBCG cluster catalog \citep{2007ApJ...660..239K} is a
cluster sample constructed from $7\,500$ deg$^2$ of photometric data
in the SDSS. To detect candidate clusters, the method identifies 
a local overdensity of cluster members along the \emph{red-sequence}.
The \emph{richness} $N_{200}$ is defined as the number of galaxy members 
which are brighter than $0.4 L_{\star}$ and lying on 
the red-sequence inside the radius of $r_{200}$,
within which the mean interior density is 200 
times the critical density of the universe.
The public maxBCG catalog consists of cluster candidates with 10 members or more.
Since the maxBCG sample was constructed from an earlier SDSS
data release, the probed area does not fully overlap with the final
SDSS coverage in the Galactic south cap, where BOSS galaxies have been
targeted. Therefore, not all of the $3\,000$
deg$^2$ of BOSS data available to date overlap with
the maxBCG cluster sample. Keeping all galaxy-cluster pairs within one
deg, the subsample used in this study includes
$6\,308$ clusters. 

Second, more recently, \citet{2010ApJS..191..254H} proposed a slightly different
optical-based detection method, the Gaussian Mixture Brightest Cluster
Galaxy (GMBCG), which relies on both the presence of the brightest cluster
galaxy (BCG) and the red-sequence to identify clusters.
Among the differences with maxBCG cluster algorithm finder is
the way to estimate the cluster redshift, where for
GMBCG it is estimated solely from the BCG photometric redshift (or
spectroscopic redshift when available).
It is claimed by \citet{2010ApJS..191..254H} that the GMBCG catalog is volume limited out to
$z=0.4$; however, to ensure no spurious correlation 
between false cluster detection and background sources,
secure foreground-background sample separation is necessary.
For this reason we limit the GMBCG sample to
the range $0.1 < z< 0.3$, taking into account the 
typical photometric redshift error.
This redshift selection, in conjunction with rejection of clusters lying outside the BOSS area,
leads to a total of $4\,631$ clusters with richness $N_{200}$ greater than 10.
Although an improved \emph{weighted} richness 
estimator is available, we employ the scaled richness estimator, which
is defined in the same way as for the maxBCG catalog.

Third, an alternative method of \citet{2008ApJ...676..868D}
does not rely on the red-sequence, but on the peak locations 
in the likelihood map generated from the convolution of 
the galaxy distribution in redshift space with aperture matched filters. 
These filters are constructed from the assumed cluster density 
profile and galaxy luminosity functions. 
This method has been applied to the SDSS \citep{2011ApJ...736...21S}, 
and the overlap with BOSS allows us to use $5\,646$ galaxy clusters
below redshift 0.3.
A noticeable feature of this catalog is
that the richness is computed as the sum of the luminosity of all
galaxies above $0.4 L_{\star}$ divided by $L_{\star}$, and is
therefore not directly comparable to
the definition used by the two previous data samples. The publicly
available version includes all clusters with an estimated richness
$\Lambda_{200}$ of 20 or higher. In the rest of the paper, we will
refer to this sample as ``AMF''.

Fourth, we use the catalog produced by \citet{2012ApJS..199...34W}.
In this method, a friend-of-friend algorithm is applied to  
luminous galaxies using a linking length of 0.5~Mpc in the
transverse separation and a photometric redshift difference within 
$3\sigma$ along the line-of-sight direction.
The center is assumed to be the position of the BCG, identified 
from a global BCG sample, as the brightest galaxy 
physically linked to the candidate cluster.
In a similar fashion to the AMF catalog, 
the richness
is computed from the total luminosity of all galaxies 
brighter than $0.4 L_{\star}$, in units of $L_{\star}$.
The sample is limited to a richness threshold of 12.
The algorithm was applied to the latest SDSS-III data
release, resulting in an increase of factor two in area. With a
larger overlap with BOSS, especially in the south Galactic region, we
obtain $15\,112$ clusters within one degree off the edge of the background
BOSS sample. In the following, this sample will be referred to as ``WHL12'',
and the richness denoted $\Lambda_{200}$ for consistency with the
notation adopted for the AMF catalog.

Finally, we complete our set of foreground cluster catalogs with the
meta-catalogue of X-ray detected clusters (MCXC)
 assembled by \citet{2011A&A...534A.109P}. This data set is a
compilation of X-ray detected clusters from a number of publicly
available data collected by the ROSAT satellite. The sample extends
to redshift $\sim0.6$ with a wide range of masses. Here we relax the
maximum redshift range to $z=0.35$ as the majority of clusters have secure
spectroscopic redshift measurements and do not suffer from false
detection caused by projection effects. The number of clusters 
in common over the DR9 (see below) released BOSS area is currently $158$.

\subsection{The Source Sample}

Our background sample is extracted from the first data release ``DR9''
of the BOSS spectroscopic survey.  BOSS aims at measuring the scale of
baryon acoustic oscillations at redshift $z=0.5$ as a sensitive
cosmological test.
The first data release, covering
$3\,000$ deg$^2$, was made publicly available in summer 2012 
through the SDSS III/BOSS DR9 data 
release\footnote{\url{http://www.sdss3.org/dr9/data\_access/}}.

We use the ``CMASS'' spectroscopic sample for which the targets 
were selected in the range $17.5 < i_{\rm AB} < 19.9$,
using color selection techniques to ensure homogeneous sample
in mass between redshifts $0.43$ and $0.7$. 
We refer to \citet{2012arXiv1207.6114M}  for a complete 
study of the galaxy stellar mass function and the 
verification of the homogeneity of the data across the full redshift range.

The BOSS DR9 data release comprises about $3\,000$
deg$^2$ in the south and north galactic areas.
We further select only primary spectroscopic galaxies,
and remove all galaxies with uncertain redshift measurements,
by imposing the flag \texttt{zwarning=0}. The total number of galaxies
used as background sources is $316\,220$.

\begin{figure*}	
  \begin{center}
    \includegraphics[width=0.99\textwidth]{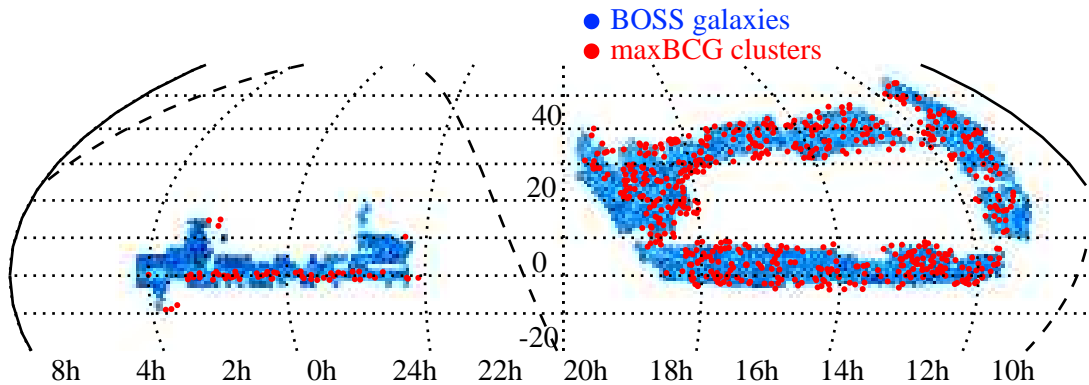} 
    \caption{Data coverage of BOSS galaxies (blue) and maxBCG
      clusters (red) in equatorial coordinates. We only display the
      clusters selected to be within one deg of BOSS area used in this study.
      The dashed line is the Galactic equatorial plane.}
    \label{fig:layout}	
  \end{center}
\end{figure*}

Fig. \ref{fig:layout} illustrates the data coverage used in this
study. The foreground cluster sample is shown in blue and the
background BOSS galaxy sample in red. We show the maxBCG clusters as
an example, and note that the WHL12 sample has a better overlap with
BOSS in the south Galactic part. The two other catalogues have a
similar coverage compared to maxBCG, and the MCXC has a full coverage
over the BOSS area but is not spatially homogeneous because it is drawn
from several independent X-ray surveys. We summarize in Table~\ref{tab:samples}
cluster sample properties.

\section{Model}

In this section we outline the formalism that describes the systematic
effect of lens magnification modifying the source selection function, 
by foreground galaxy clusters. Lensing magnification is caused by both
isotropic and anisotropic focusing of light rays due to the presence
of massive foreground objects acting as gravitational lenses.  The
former effect is described by the convergence,
$\kappa(r)=\Sigma(r)/\Sigma_{\rm crit}$, the projected mass density
$\Sigma(r)$ in units of the critical surface mass density,
\begin{equation}
\Sigma_{\rm crit} = \frac{c^2}{4\pi G}\frac{D_{s}}{D_l D_{ls}} \, ,
\end{equation}
where $D_l$, $D_s$, and $D_{ls}$ are the proper angular diameter
distances from the observer to the lens, the observer
to the source, and the lens to the source.
The latter effect is due to the gravitational shear $\gamma(r)=\gamma_1+i\gamma_2$
with spin-2 rotational symmetry \cite[see][]{2001PhR...340..291B}.

Since gravitational lensing conserves surface brightness, the apparent
flux of background sources increases in proportion to the
magnification factor. This shift in magnitude implies that the
limiting luminosity at any background redshift lies effectively 
at a fainter limit given by $L_{\rm lim}(z)/\mu(z)$, hence
increasing the surface density of magnified sources behind foreground
lenses.  On the other hand, the number of background sources per unit
area decreases due to the expansion of sky area.  These two effects
compete with each other, and the effective variation in the source
number density $n_{\rm eff}$, known as the magnification bias
(BTP95),
depends on the steepness of the source number
counts as a function of the flux limit $F$.

For background sources at redshift $z$, the magnification bias is expressed as
\begin{equation}
  \label{eq:magbias}
  n_{\rm eff} \left [ > F(z) \right ]= 
  \frac{1}{\mu(z)} n_0 \left [ >  \frac{F(z)}{\mu(z)} \right ] \, ,
\end{equation}
where $n_0$ is the unlensed number density of background sources, 
$L$ the limiting luminosity of the background sample, 
and $\mu$ the magnification,
\begin{equation}
\label{eq:mu}
\mu =\frac{1}{(1-\kappa)^2-|\gamma|^2}.
\end{equation}

For the case of weak gravitational lensing,
the magnitude shift induced by
magnification is sufficiently small, so that the source number count 
can be locally approximated by a power law at the limiting luminosity.  
This simplifies Eq.~\ref{eq:magbias} to 
\begin{equation}
  \label{eq:neff}
  n_{\rm eff} (z) = \mu^{\beta(z,L) - 1} n_0(z) \, ,
\end{equation}
with $\beta$ the logarithmic 
slope of the luminosity function $\Phi$ evaluated at the limiting luminosity: 
\begin{equation}
\label{eq:meanz}
  \beta(z,L) = - \frac{\ud \ln \Phi \left[ L', z \right ]}{\ud \ln  L'}
  \bigg |_{L}\, .
\end{equation}
We retrieve the well-known result that if the count slope $\beta(z,L)$
is greater than unity,  the net effect of magnification bias is to increase
the source number density, or decrease otherwise (BTP95). 
If the count slope is unity, the net magnification effect on the source counts
vanishes.  
In the strict weak-lensing limit, the magnification bias is directly
related to the projected mass distribution as
$n_{\rm eff}(z)/n_0(z)\approx 2[\beta(z,L)-1]\kappa(z)$.\footnote{In the present analysis, we have used the full expression without the weak-lensing approximation.}

The limiting magnitude and the count slope 
vary with redshift, so that the integrated magnification-bias 
effect will translate into an enhancement in mean source redshift as
\begin{equation}
  \bar{z}_{\rm lensed} = \frac{\int n_{\rm eff} (z) \, z \,
    \ud z}{\int n_{\rm eff} (z) \,  \ud z} \, .
\end{equation}

The limiting luminosity can be computed from the apparent limiting
magnitude for a given survey as
\begin{equation}
  -2.5 \log_{10} L(z) = i_{\rm AB} - 5\log_{10} \frac{d_L(z)}{10\,{\rm pc}} - K(z)
\end{equation}
where $i_{\rm AB} = 19.9$ is the limiting magnitude of BOSS, 
$d_L$ the luminosity distance, and $K$, the $K$-correction:
\begin{equation}
\label{eq:kcorr}
  K(z) = 2.5(1+z) + 2.5\log_{10} \left [\frac{L(\lambda_e)}{L(\lambda_0)}
\right ] \, .
\end{equation}

In our analysis, we restrict the source redshift range to $0.43 < z < 0.7$, 
where the BOSS target selection is established to be uniform \citep{2013AJ....145...10D}. 
We assume a \citet{1976ApJ...203..297S} luminosity function with the parameters
measured in the VIMOS VLT Deep Survey
\citep[VVDS,][]{2005A&A...439..845L} 
by \citet{2005A&A...439..863I}. The authors provide a detailed
measurement of the luminosity function in all $UBVRI$ optical bands of
the survey over a wide redshift range, $0.2<z<2.0$. This survey has
the advantage of being considerably deeper than the BOSS survey, with redshift
completeness to a fainter limiting magnitude and therefore describes
better the underlying luminosity function, in terms of the slope as a
function of redshift and magnitude required here for our lensing
predictions.
To minimize the uncertainties on the $K$-correction, we take the
rest-frame $V$-band luminosity function measured at redshift
$\sim0.5$, which best matches the redshifted galaxies observed in the
$i-$band at redshift 0.5, so that this choice allows us to neglect the
second term in Eq.~\ref{eq:kcorr}.

Finally, to account for luminosity evolution as function of redshift,
we employed the parametrization of \citet{2007ApJ...665..265F}, such
that our assumed Schechter parameters are:
\begin{eqnarray}
  M_{*}  & = & -22.27 - 1.23\times(z-0.5) \\
  \alpha  &  =  & -1.35 \, .
\end{eqnarray}
Note that no precise normalization of the luminosity function is
required for our analysis but only the gradient of the logarithmic slope
$\beta$ at the limiting luminosity as a function of background
redshift, which has been
well determined from the VVDS, and confirmed 
by other deep probes of similar volume such as the COMBO-17
 photometric survey \citep{2003A&A...401...73W} or 
DEEP2 spectroscopic survey \citep{2006ApJ...647..853W}.
In particular, the $M_{\star}$ parameter measured in the $B-$band
is found to be in excellent agreement among those three 
deep surveys at $z=0.5$ \citep[][Fig.~7]{2007ApJ...665..265F}.

The VVDS field-of-view is one~deg$^2$, comprising $11\,034$
 redshift measurements, allowing us to determine the respective levels
of Poisson uncertainty and cosmic variance into the Schechter
parameters we require for our predictions.
We quantify in Sec.~\ref{sec:err-model} the impact of these
assumptions on the size of the mean redshift depth enhancement
predicted. In particular we show that these model uncertainties
arising from the parameterisation of the luminosity function are
substantial compared to the detected signal, 
however smaller than our statistical errors on the measurements
from the current DR9 release.

To date, a number of studies
\citep{2011ApJ...729..127U,2011ApJ...733L..30H,2012ApJ...754..143F}
have measured a magnification-bias signal by assuming an effective
single-plane source redshift for a given background sample and
comparing their number density to a random sample. Here we do not need to approximate
the depth as we have precise spectroscopic redshift
measurements for all individual sources, providing a direct estimate for
the enhancement of mean source redshift behind clusters with respect
to that of the total sample.

Our method developed here has two main advantages over the standard
magnification-bias because we are simply measuring an averaged depth
enhancement of the background galaxies, rather than the lensing
induced change in the surface density of galaxies. The latter effect
requires careful correction for screening by foreground/cluster
galaxies. Secondly, since we probe a
relatively narrow redshift range of galaxies having a well-understood
selection function than for typical faint background source counts used in the
magnification bias work located at $z\sim 1-3$.  Therefore, our method enables
in practice a relatively better understood determination of the statistical lensing
effect for the foreground clusters we are examining.

To quantify and characterize the cluster mass distribution,
we compare the observed cluster lensing profiles with 
analytic spherical halo density profiles.  
In the present study, we consider (1) the 
\citet[][hereafter NFW]{1997ApJ...490..493N}
and (2) the singular isothermal sphere (SIS) models.
The former is a theoretically- and observationally-motivated 
model of the internal structure of cluster-sized halos
\citep[e.g.][]{2010PASJ...62..811O,2011ApJ...738...41U},
$\rho(r)\propto (r/r_s)^{-1}(1+r/r_s)^{-2}$ with $r_s$
the characteristic radius at which the logarithmic density slope is isothermal.
The latter model provides a simple, one-parameter description of 
isothermal density profiles, $\rho(r)\propto r^{-2}$.

The two-parameter NFW model can be specified by
the degree of concentration, $c_{200}=r_{200}/r_s$, and the halo mass, 
\begin{equation}
  M_{200} = \frac{4\pi}{3} \, 200 \, \rho_{\rm crit} (z) \,
  r_{200}^{3},
\end{equation}
the total mass enclosed within a
sphere of radius $r_{200}$, within which the mean interior density is
200 times the critical mass density at the cluster redshift, $\rho_{\rm crit}(z)$.
Here we employ the mean concentration-mass
relation of \citet{2013ApJ...766...32B},
derived from $\Lambda$CDM cosmological $N$-body simulations 
covering a wide halo mass range of $2\times 10^{12}$--$2\times 10^{15}M_\odot\,h^{-1}$
and a wide redshift range of $z=0$--$2$.  We use their fitting formula for the full-sample relation
$c_{200}(M_{200},z)$. The NFW profile is thus specified by $M_{200}$ alone.
The use of the \citet{2013ApJ...766...32B} relation is motivated by recent detailed cluster
lensing work
\citep{2012ApJ...757...22C,2012ApJ...755...56U,2013arXiv1302.2728O} finding a good agreement
with their predictions for high-mass clusters ($M_{200}\sim 10^{15}M_\odot$) at $z=0.2$--$0.4$.
We employ the radial dependence of the projected NFW lensing profiles
given by \citet{2000ApJ...534...34W}.

For the SIS model, the magnification is given by
\begin{equation}
   \mu_{\rm SIS} (\theta) = \frac{1}{1 - \theta_E/\theta} \, ,
\end{equation}
with $\theta_E=4\pi (\sigma_v/c)^2 D_{ls}/D_s$ the Einstein radius.
Here, $\sigma_v$ is the one-dimensional velocity dispersion related to the halo mass,
\begin{equation}
  \sigma_v = \left [ \frac{\pi}{6} 200 \rho_{\rm crit} (z) M^2_{200}
G^3 \right ]^{1/6} \, .
\end{equation}

Finally, the magnification factor is plugged into Eqs.~\ref{eq:neff} and
\ref{eq:meanz}, so that the depth magnification is computed as a function of the physical distance
at the cluster redshift as
\begin{equation}
  \label{eq:depth_mag}
  \delta_z  (r) = \frac{\bar{z} (r)}{ \bar{z}_{\rm total}} - 1  \, .
\end{equation}

\begin{figure}	
  \begin{center}
    \begin{tabular}{c@{}}
      \includegraphics[width=0.5\textwidth]{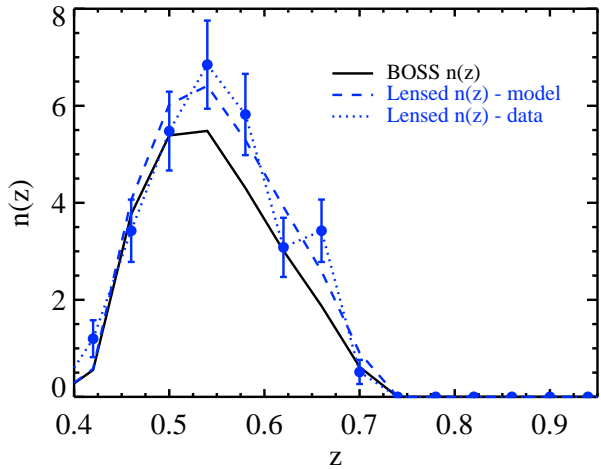}
    \end{tabular}
    \caption{Redshift distribution of BOSS background galaxies
      lensed by foreground SDSS clusters 
      (in blue: dashed line for our model; 
       dotted line with error bars for the AMF measurements) 
      compared to the unlensed distribution
      (black solid line), and averaged within $r=0.2$\,Mpc from the cluster center.
      The model assumes an effective cluster mass of $M_{200} = 1.81\times 10^{14} M_{\odot}$ 
      at $z=0.2$.
      The observed enhancement in mean source redshift is $\delta z \sim 0.01$.}
    \label{fig:results_nz}	
  \end{center}
\end{figure}

The total effect of magnification on the redshift distribution $n(z)$
of BOSS source galaxies is shown in Fig.~\ref{fig:results_nz}.

\section{Measurements}
\label{sec:measurements}

Given the sparse source density around foreground clusters, 
we must measure the signal around stacked clusters to increase the 
signal-to-noise ratio.

To ease the computation of the mean redshift of BOSS galaxies 
around clusters,  we employed a modified version
of the \citet{1983ApJ...267..465D} two-point cross-correlation estimator:
\begin{equation}
  \label{eq:peebles}
  w(r) = A \,\frac{L  S}{L R} -1 	
\end{equation}	
where $L$ is the lens cluster sample, $S$ the source BOSS galaxy sample
weighted by their redshift and $R$ the un-weighted BOSS galaxy
sample. Pairs are computed at the redshift of the cluster as function
of physical scale.

We can write the previous equation explicitly:
\begin{eqnarray}
  L  S (r) & = &\sum\limits_{i=1, j=1}^{n_{\rm l}(r), n_{\rm s}(r)} 1 \times z_j \\
  L  R (r)    & = & n_{\rm l}(r) \times n_{\rm s}(r)
\end{eqnarray}
so that
\begin{equation}
  \frac{L  S}{L  R}(r)  =   \bar{z} (r)
\end{equation}
and
\begin{equation}
  A^{-1}  = \frac{1}{N_{\rm l}}\sum\limits_{i=1, j=1}^{N_{\rm l}, N_{\rm s}} 1
    \times z_j = \bar{z}_{\rm total} \, ,
\end{equation}
where the subscripts $l$ and $s$ 
denote the lens and source samples, respectively. 
Equation~\ref{eq:peebles} then reduces to
\begin{equation}
  \label{eq:peebles_modified}
  w(r) = \frac{\bar{z} (r)}{\bar{z}_{\rm total}} - 1 \equiv \delta_z  (r) \, ,
\end{equation}
as defined by Eq.~\ref{eq:depth_mag}.

In practice we perform the measurements using the software
{\sc Swot} \citep{2012A&A...542A...5C}, a fast two-point correlation code optimized
for large data sets. The algorithm makes use of large 
scale approximations, tree-code structured data, and parallel computing to
considerably accelerate pair counting. 
Correlating $\sim300\,000$ background objects with $\sim6\,000$
foreground objects takes about five minutes on a desktop computer.

Error bars are estimated by generating foreground cluster positions over the
BOSS area: using $20\,000$ points with random projected positions on the sky and a 
volume-weighted random redshift in the range $0.1 < z < 0.3$ , we compute $w(r)$ in a 
similar fashion as for the data samples, using the same binning. We repeat 
the process 100 times and compute the dispersion of the ensemble. Error bars
for the signal of each cluster sample are further multiplied by $\sqrt{20\,000/N_{\rm clusters}}$ 
to account for Poisson error. Additionally, the significance of the 
detection for each sample is computed using the rescaled 
covariance matrix.

For optically detected cluster samples, 
we compare our measurements to the expected theoretical 
signal by first converting the richness to the mass,
using the maxBCG mass-richness relationship calibrated from 
X-ray and weak lensing 
by \citet{2009ApJ...699..768R}:
\begin{equation}
  \label{eq:mr-rozo}
  \frac{M_{500}} {10^{14} M_{\odot}} = \exp(0.95)\left
    (\frac{N_{200}}{40} \right )^{1.06} \, ,
\end{equation}
and we further convert $M_{500}$ into $M_{200}$ using the method described in
\citet{2003ApJ...584..702H}, and the concentration-mass relationship of
\citet{2013ApJ...766...32B}. We employ the same relation for 
GMBCG, as the richness definition is identical to maxBCG.

As for the WHL12 sample, the authors chose a different
definition for the richness (see Sec.~\ref{sec:data-lens}),
but also provide an independent X-ray/lensing calibrated 
mass-richness relationship \citep{2010MNRAS.407..533W}:
\begin{equation}
  \label{eq:mr-wen}
 	 \frac{M_{200}} {10^{14} M_{\odot}} = 10^{-1.49}
         \Lambda_{200}^{1.17} \, .
\end{equation}
Since the richness definition assumed for AMF is similar to that of WHL12
we use the same parametrisation for these two samples.

For the X-ray cluster sample, an estimation of the $M_{500}$ 
mass is directly provided, however due to the low angular resolution
of ROSAT, X-ray mass estimates may be underestimated. To correct
for this, we matched the MCXC clusters with the local cluster sample presented
in \citet{2009ApJ...692.1033V}, and compared the value of
$M_{500}$:
\begin{equation}
  M_{500}^{'} =  10^{-0.951} M_{500} ^{1.067} \, .
\end{equation} 
Then we simply convert $M_{500}^{'}$ into $M_{200}$ as described above.
It is found that the MCXC-based value is
about 25\% smaller than those derived by \citet{2009ApJ...692.1033V}
at the effective mass of our subsample $M_{200}\sim5\times10^{14} M_{\odot}$.

Finally to perform a strict comparison between the stacked measured signal and
the prediction, we compute the composite halo-mass theoretical signal:
\begin{equation}
  \label{eq:mr-wen}
\delta_z (r) = \frac{1}{N_{\rm l}} \sum\limits_{i=1}^{N_{\rm l}}
  \delta_{z,i} (r, M_{200}) \, .
\end{equation}

\section{Results}
\label{sec:results}

\begin{figure*}
  \begin{center}
    \includegraphics[width=0.49\textwidth]{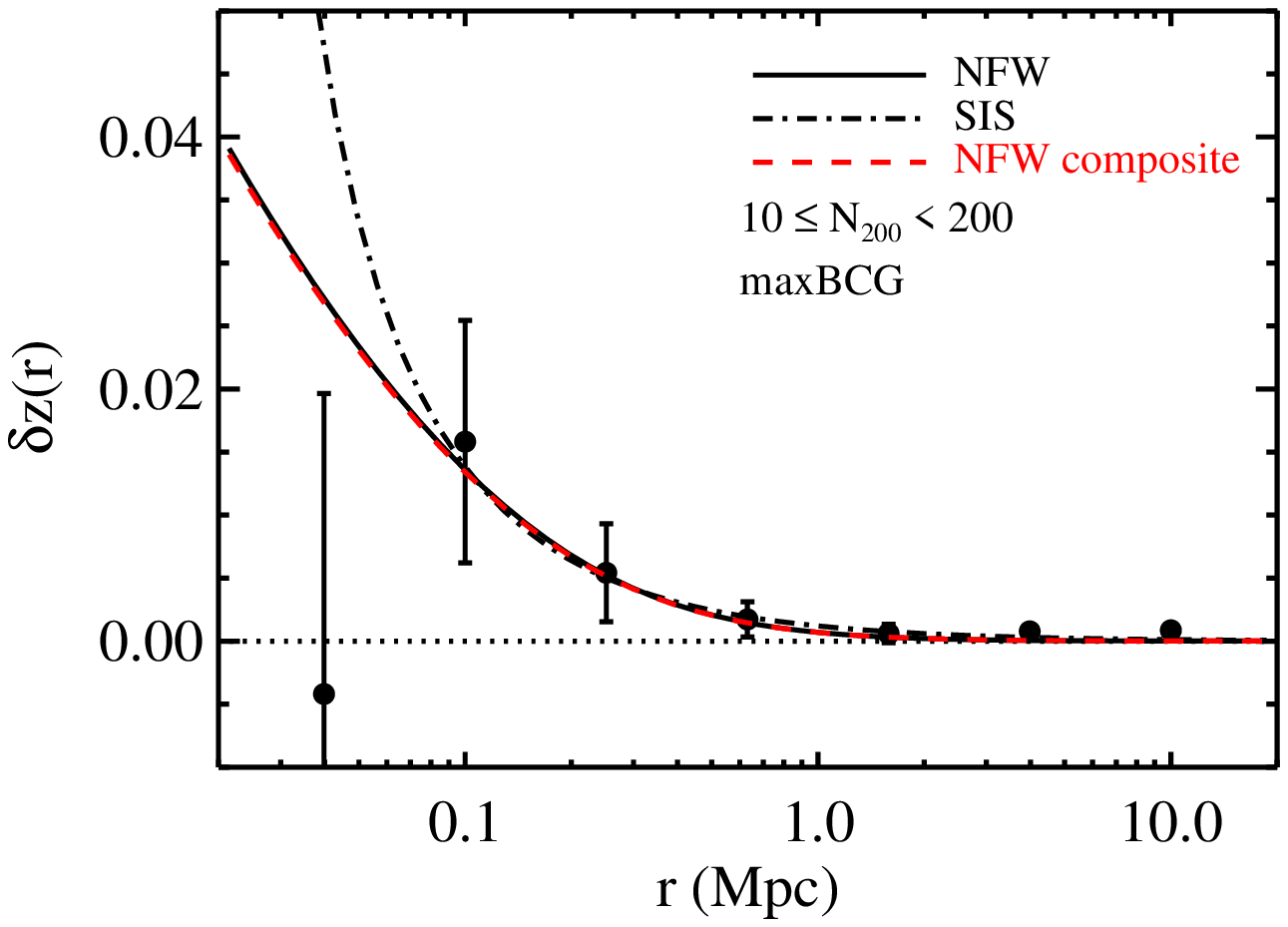}
    \includegraphics[width=0.49\textwidth]{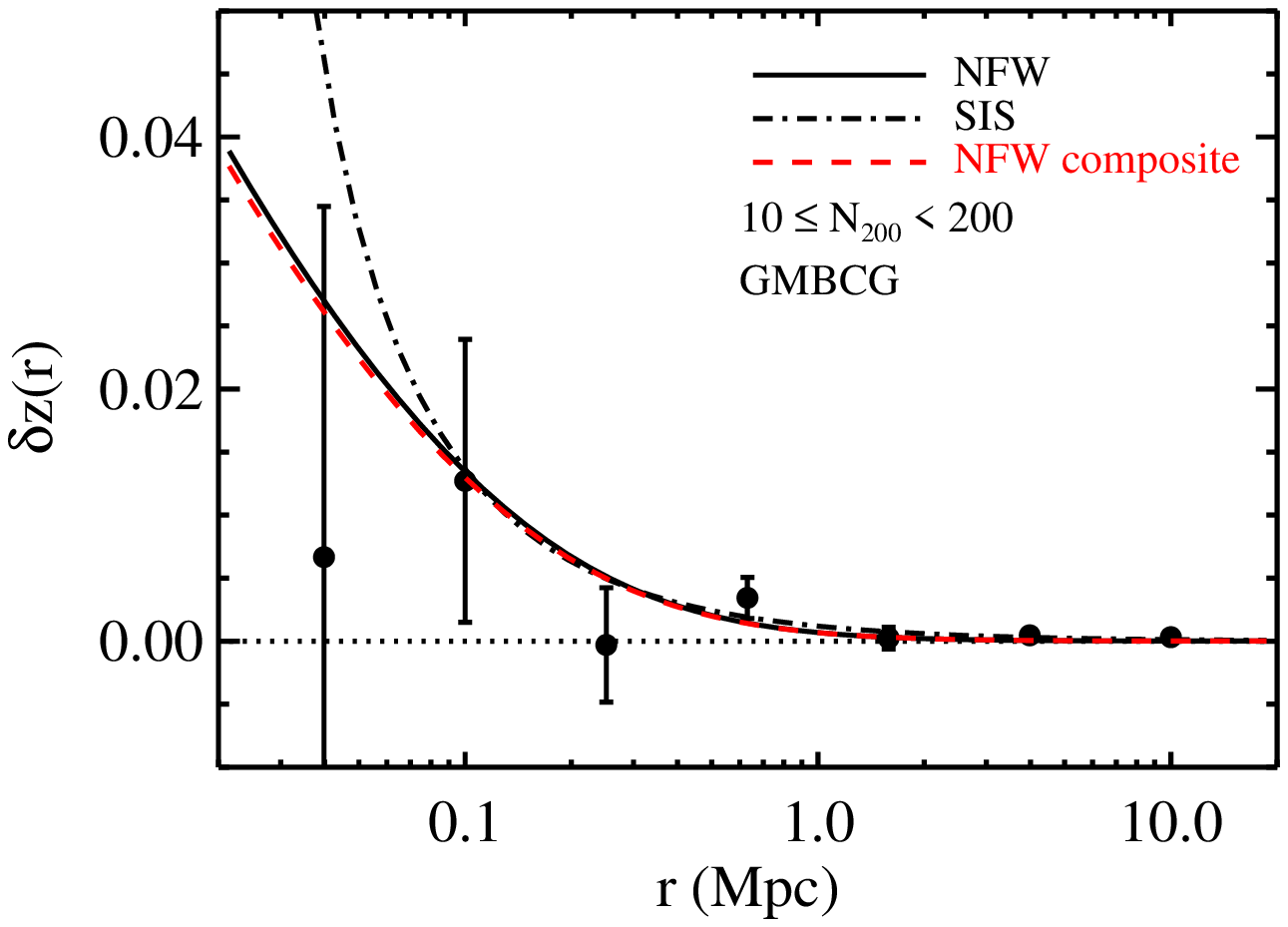} 
    \includegraphics[width=0.49\textwidth]{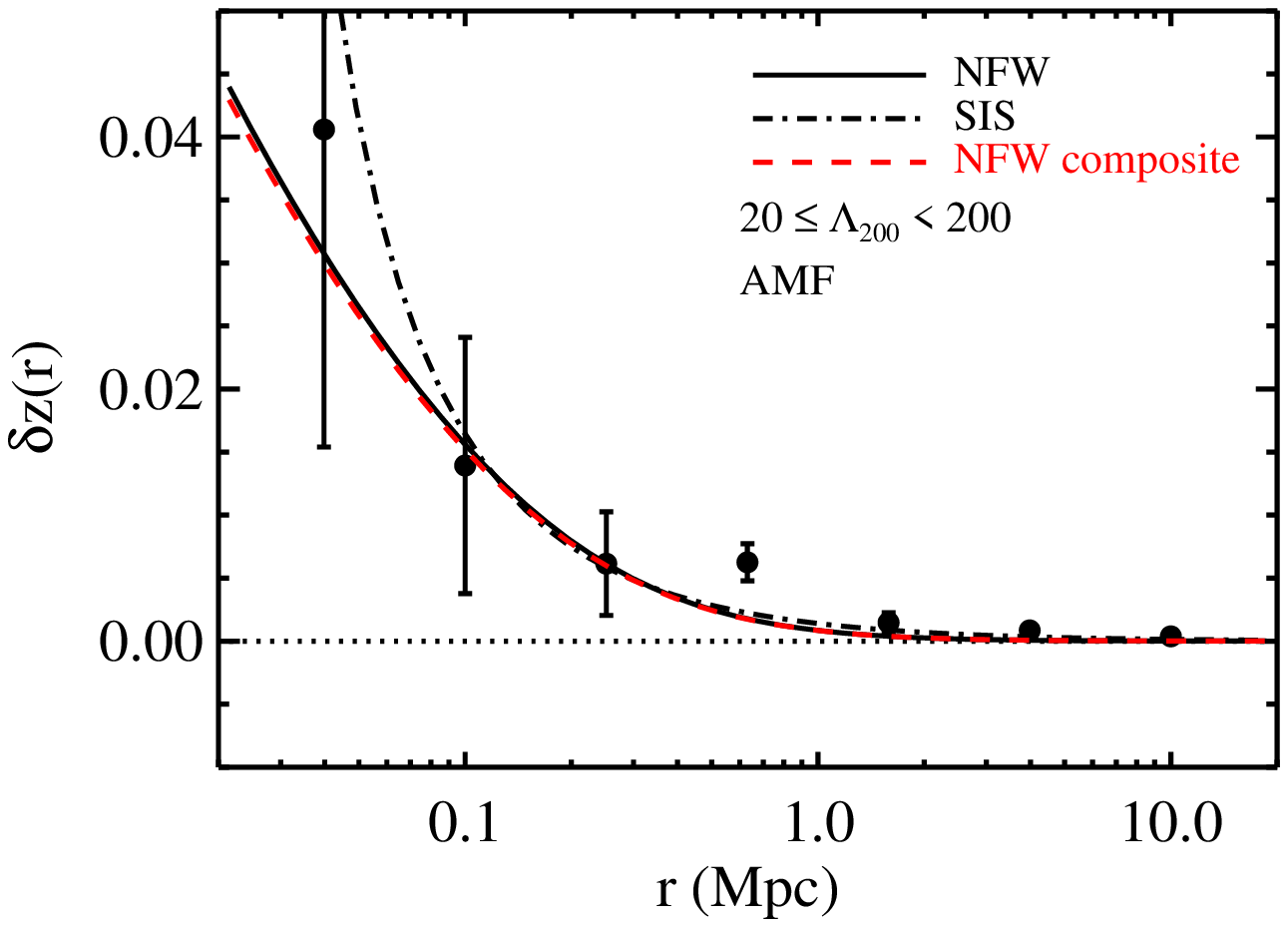} 
    \includegraphics[width=0.49\textwidth]{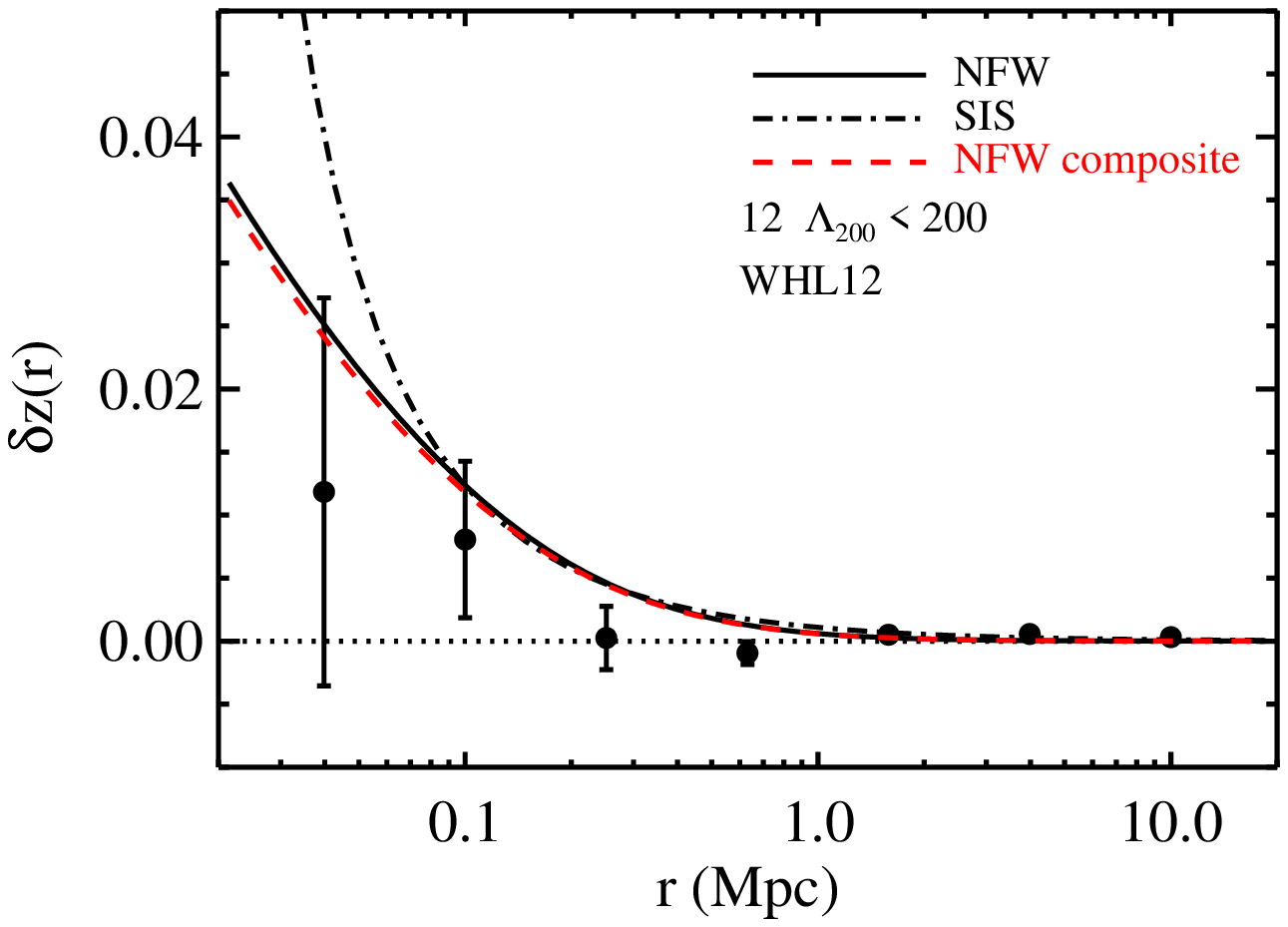}
    \caption{Mean redshift increase of BOSS galaxies measured around
      SDSS clusters from the maxBCG (top left), GMBCG (top right), AMF (bottom left), and
      WHL12 (bottom right) samples. The red dashed (composite) and black
      solid lines are NFW profile predictions assuming a mass-richness
      relationship calibrated from X-ray and weak lensing
      measurements taken from the literature. We also show the SIS
      profile prediction (dotted dashed line). Error bars were computed by calculating the 
       background mean redshifts radially around 20,000 randomized foreground
      positions over the BOSS area and normalised to the cluster sample size.}
    \label{fig:results-SDSS}	
  \end{center}
\end{figure*}

\begin{figure}
  \begin{center}
    \includegraphics[width=0.49\textwidth]{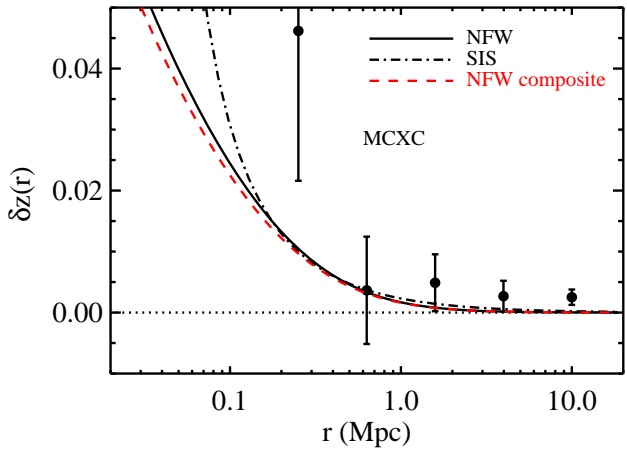}
    \caption{Same as Fig.~\ref{fig:results-SDSS}, but for the X-ray
      MCXC sample.}
    \label{fig:results-X-ray}	
  \end{center}
\end{figure}

We show in Fig.~\ref{fig:results-SDSS} our measurements for 
the four maxBCG, GMBCG, AMF, and WHL12 SDSS cluster
samples.  Similarly, Fig.~\ref{fig:results-X-ray} shows our results 
for the X-ray MCXC cluster sample. 

\begin{figure}	
  \begin{center}
      \includegraphics[width=0.49\textwidth]{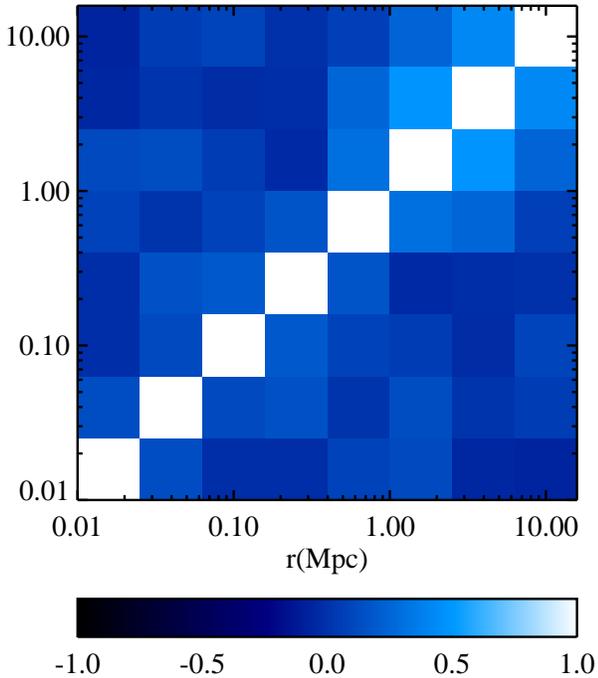}
    \caption{Correlation matrix of the covariance matrix estimated
      from the 100 subsamples with randomized cluster positions 
      in the BOSS area. The result shown here was computed using 
      $20,000$ random points for each of the 100 subsamples.}
    \label{fig:corr_matrix}	
  \end{center}
\end{figure}

As described in Sec.~\ref{sec:measurements}, the errors are obtained
from the standard deviation of 100 subsamples with $20\,000$ 
randomized cluster positions over the BOSS area, and rescaled
for each individual sample depending on the number of clusters 
to account for Poisson uncertainty.
The detection significance, shown in Table~\ref{tab:samples} 
for each individual sample, is computed for 
all data points with 10 cluster-galaxy pairs or more, using
the covariance matrix of the random samples 
up to $r  = 10.0$\,Mpc.
The resulting covariance matrix is displayed in Fig.~\ref{fig:corr_matrix}:
the higher correlation at large scales translates 
into the fact that the same background galaxy may
be used for several cluster-galaxy pairs. On the other hand,
there is little correlation at smaller scales ($<1$\,Mpc)
due to the relatively sparse distribution of foreground clusters (approximately, two per sq. deg).

\begin{deluxetable*}{l r c c c r r}
  \centering
  \tabletypesize{\scriptsize}	
  \tablecaption{Properties of cluster samples. \label{tab:samples}}
  \tablehead{
    \colhead{Sample} & \colhead{$N_{\rm clusters}$} &
    \colhead{Effective $M_{200}$ mass $[\times10^{14} M_{\odot}]$}  &
    \colhead{Redshift} & \colhead{Detection method} &	
    \colhead{Significance ($\sigma$)}& \colhead{Reference}
  }
  \startdata
  maxBCG & $6\,308$ & 1.48 & $0.1 < z < 0.3$ &red sequence & 4.8 & 1 \\ 
  GMBCG    & $4\,631$  &   1.40  & $0.1 < z < 0.3$ & red sequence + BCG& 2.8 & 2\\ 
  AMF       & $5\,646$   & 1.81 & $0.1 < z < 0.3$ & aperture matched filter &
  4.9 & 3\\ 
  WHJ12     &  $15\,112$  & 1.18 & $0.1 < z < 0.3$ & photo-z &  3.5 & 4\\ 
  MCXC    &  $158$   & 5.00  & $0.1 < z < 0.35$ & X-ray &  2.8 & 5
  \enddata
  \tablerefs{(1) \citet{2007ApJ...660..239K}; (2)
    \citet{2010ApJS..191..254H}; (3) \citet{2011ApJ...736...21S}; (4) 
    \citet{2012ApJS..199...34W} ; (5)\citet{2011A&A...534A.109P}.} 
\end{deluxetable*}

For each case, the red-dashed line represents the composite-NFW
lensing signal expected for the respective cluster sample, obtained
assuming a mass-richness relationship calibrated from X-ray and
weak-lensing measurements taken from the published literature: given
by \citet{2009ApJ...699..768R} for maxBCG and GMBG
(Eq.~\ref{eq:mr-rozo}) and by \citet{2012ApJS..199...34W} for AMF and
WHL12 (Eq.~\ref{eq:mr-wen}), whereas the $M_{200}$ masses for MCXC
have been translated from their $M_{500}$ values assuming the NFW
form.

For comparison, we also plot the expected lensing signal obtained assuming 
a single effective mass $\langle M_{200}\rangle$,
for the NFW (black solid line) and SIS (black-dotted dashed line) profiles. 
At small scales, there is a small difference between the composite versus single
NFW profiles, which however is not significant due to the large statistical errors.
As for the SIS case, overall, 
it is likely that the lensing signal at small scales is overestimated.
Here, it is reassuring to see that the NFW model provides a better description of
the observed lensing signals because the lensing-based mass-richness
relationships were calibrated assuming the NFW form in all cases for the maxBCG relation
\citep{2007ApJ...656...27J, 2008JCAP...08..006M, 2009ApJ...703.2217S},
and most cases for the WHL12 relation 
\citep[][and references therein]{2010MNRAS.407..533W}.

These results demonstrate the robustness of our measurements and
appear to be in excellent agreement with previous lensing studies.
Although the large statistical errors prevent us from accurately
testing both the amplitude and slope of the mass-richness relationship
for the current BOSS sample, it is interesting to compare the results
between the different cluster samples.  With a detection significance at the
nearly 5$\sigma$ level, the maxBCG and AMF samples seem to give the
highest-confidence detections among our cluster samples.
The AMF catalog catalogue has the highest minimum richness cut ($\Lambda_{200} \ge 20$) 
but the averaged signal appears to be relatively higher so that 
the lower number of clusters above this richness cut is 
almost compensated by a greater lensing signal due to their higher
cluster masses.

On the other hand, 
the measured lensing signal in the WHL12 sample is shown to be 
relatively low compared to the model prediction based on the same mass-richness relationship,
despite its larger sample size and greater overlap with current BOSS sky area.
This could be a consequence of a relatively higher impurity of the cluster sample,
due perhaps to a higher rate of false detections, which will lead to an underestimation
of the geometric lensing signal as found here.
Similarly, we obtained for GMBCG a relatively low level of detection significance,
$2.4\sigma$.
It is interesting to note that  \citet{2012ApJS..199...34W} find a lower 
matching rate of their WHL12 clusters with GMBCG, 
compared to that with maxBCG.
This could indicate that these two samples have a higher level of 
contamination by false cluster detections. 
Another interesting point is that
the cluster redshift scatter between 
WHL12 and GMBCG can be as high as several times the photometric redshift
errors, as shown by their comparison \citep[][Table 8]{2012ApJS..199...34W}.
A more detailed and careful analysis would be certainly required to confirm such a conclusion.
We shall come back to this issue and its consequence for
magnification bias contamination in Sec.~\ref{sec:err-sys}.

The MCXC X-ray sample has the largest statistical 
errors due to the small number, 158, of clusters (Fig.~\ref{fig:results-X-ray}).
However, it is worth noting that 
there is a likely excess of the lensing signal at $r\simeq 2$--10\,Mpc ($>r_{200}$)
with respect to the NFW predictions,
which cannot be explained by simply increasing the halo masses.
In the context of $\Lambda$CDM, this large-scale excess signal can be naturally explained 
by the two-halo term contribution 
due to large-scale structure associated with the central clusters \citep[see][]{2011MNRAS.414.1851O}.
We note that, in contrast to the gravitational shear, 
magnification is sensitive to the sheet-like mass distribution, and therefore can be used 
as a powerful tool to probe the two-halo term in projection space.

\section{Error Analysis}
\label{sec:err}

\subsection{Sources Sample Variance}
\label{sec:error-stats}

\begin{figure}	
  \begin{center}
    \begin{tabular}{c@{}}
      \includegraphics[width=0.5\textwidth]{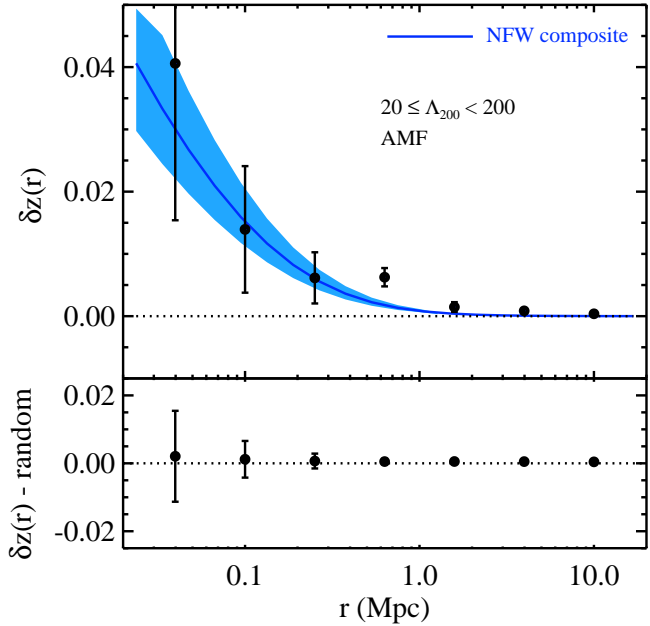}
    \end{tabular}
    \caption{Measured signal for the AMF sample (top) compared to the
      mean signal around random positions drawn in the BOSS area (bottom). The
      error bars show the dispersion for 100 samples with randomised cluster
      positions, mainly accounting for the source sample variance around the cluster
      sample.  In blue we display the expected NFW signal as in Fig.\ref{fig:results-SDSS},
      where the shaded area represents an estimate on our model uncertainties due to the
      	assumptions on the luminosity function.}
    \label{fig:results_sys}	
  \end{center}
\end{figure}

As described in Sec.~\ref{sec:measurements},
our error estimate simply reflects the statistical variation of the mean redshift
of sources across the BOSS area. 
This estimator primarily provides a measure of the source 
sample variance given a sample of foreground clusters, 
because it measures the fluctuations of the number counts across the field.
We display the magnification signal as measured for AMF
(for which we find the highest level of detection significance)
in the top panel of Fig.~\ref{fig:results_sys},
compared to the mean signal from the random samples shown in the bottom panel.
The random-sample signal is consistent with null detection, indicating that 
the 100 random samples are sufficiently large for realistic estimates of the errors.

\subsection{Spectroscopic Redshift Errors}

The errors on individual source redshifts might also affect our statistical lensing measurements if
only a small number of cluster-galaxy pairs are used at small scales ($\simlt 0.05$\,Mpc). 
However, the mean spectroscopic error is $\Delta z = 1.3\times10^{-4}$, so that 
our measured signal is tow orders of magnitude larger than that. 
Hence, we conclude that this should not have a significant impact on our measurements.

\subsection{Photometric Calibration}

The primary requirement of our analysis method is 
an homogeneous target selection across the
survey field. 
In particular, a spatial variation of the photometric calibration would
affect both the cluster detections and the BOSS target
selection. 
In our analysis, we can rely on the accurate and homogeneous calibration
of the SDSS data \citep{1996AJ....111.1748F,1998AJ....116.3040G}.
We have checked that a variation of a few 0.01 magnitude would cause
a mean redshift change less than $\delta z = 5\times10^{-4}$, 
which is again significantly smaller than the measured signal.

\subsection{False Cluster Detections}
\label{sec:err-sys}

Perhaps the most important source of systematics in magnification bias
methods originates from the potential contamination by physically
associated lens-source pairs that create a spurious correlation signal
indistinguishable from the lensing signal.
Since we use spectroscopic redshift information for the background, 
low-redshift contamination by background sources can be neglected here.
Hence, the only possible source for such systematics would arise from
erroneously-detected distant clusters.
Let us consider a galaxy overdensity (such as rich clusters or filamentary structure)
located in the source range $0.43 < z < 0.7$ whose member galaxies 
were misattributed to foreground galaxies with lower photometric redshifts,
which can be incorrectly detected as a low
redshift cluster. Since cluster-sized overdensities can have a significant impact on 
the redshift distribution, the resulting mean redshift could be noticeably different
from that of the whole sample.

The maxBCG and GMBCG samples would have the lowest chance of being
affected by this effect, because their detection algorithms rely on bright red
galaxies, for which we expect a very low rate of photometric redshift
outliers (defined as being off by a few sigmas) due to a clear 
measurable Balmer break in the redshift range considered here, and the high rate of 
spectroscopic redshifts for these objects.
On the other hand, for the WHL12 and AMF samples which make use of the full
galaxy population with photometric redshifts, such misidentification is more likely
to happen.

Interestingly,  Fig.~4 in \citet{2008ApJ...674..768O} 
shows a photometric-to-spectroscopic redshift comparison of the 
\texttt{CC2} sample that was used for AMF. A closer look at the published figure
suggests that very few galaxies (one can count only six galaxies 
out of $\sim10\,000$ -- less than
0.1\%) scatter from the redshift range $[0.43:0.7]$ to $[0.1:0.3]$. 
Taking into account the increased scatter in overdense regions,
a threshold richness of $\sim10$ in cluster catalogs 
implies that only those clusters which are richer
than a few hundreds would be large enough to cause such a false
detection at low redshift.

Also rare, but more likely to occur than in the previous case, 
is massive filaments aligned with the line-of-sight, which can
lead to enhanced low-z contamination.
This might cause a false cluster detection \emph{and} also impact the
mean redshift estimation of the background sample. Again from 
Fig.~4 in \citet{2008ApJ...674..768O}, it is clearly seen 
that the scattered galaxies 
lie below the mean redshift of BOSS galaxies. This means that
the expected direction of this effect would be to decrease the mean
redshift and bias low our measurements. This could constitute 
a plausible explanation for the low signal level measured 
from WHL12, since they employ photometric redshifts in their detection
algorithm.

It is worth noting that such large overdensities would be easy to 
spot and mask out in a spectroscopic background sample,
when the statistical significance of the BOSS survey improves.

\subsection{Cluster Miscentering}

Misidentification of cluster centers is another potential source of
systematic errors for cluster lensing measurements at small scales. 
Recently, \citet{2012ApJ...757....2G} 
studied the impact of the choice for the cluster center on the measured 
lensing signal based on X-ray galaxy groups detected in the COSMOS field.
Their findings show that at small scales ($<0.1$~Mpc),
choosing the BCG over the most massive galaxy close to the X-ray peak
position (their center definition yielding the best result) reduces the signal by about
20\% within 75~Kpc. 

\citet{2007ApJ...656...27J} demonstrated that the
lensing convergence $\kappa$, 
which is locally related to the magnification to first order,
is less affected by cluster miscentering than the weak shear, and that
smoothing due to miscentering effects nearly vanishes at twice the
typical positional offset from the cluster mass centroid.  This
indicates that our results would not be affected by the miscentering
effects beyond a radius of $r=0.15$\,Mpc.  In our present analysis,
the statistical uncertainty is too large to be able to estimate the
degree of miscentering or the likely level of correction from our
data.  Still, it is worth noting that our measurements show the
lensing signal decreases at small scales for three out of the four
samples, although these cluster samples are not entirely independent
with sizeable proportions of clusters in common \citep{2012ApJS..199...34W}.

\subsection{Model Uncertainties}
\label{sec:err-model}

In this work, while we attempt to demonstrate the feasibility and great potential 
of this new method, we are limited by statistical
uncertainties and thus unable to constrain both the amplitude and slope of 
the mass-richness relationship.
When BOSS is completed, it will increase the respective size of 
background sample and the large area of sky covered -- which means we can
also increase the size of the foreground cluster samples by a factor of three --
leading to an anticipated  factor-three increase in the signal-to-noise ratio.
This will then allow us to subdivide cluster samples 
into bins of richness, thus complementing the standard shear-based mass measurements.
A further substantial improvement is expected in near future with 
bigBOSS and the NIR spectroscopic survey of EUCLID, which will increase the
sample size of background galaxies by one order of magnitude.
More importantly, EUCLID will obtain a sufficiently large number of
spectra of background sources at $z > 1$, pushing the foreground 
cluster sample to higher redshift than shear measurement will allow.
Similarly, the ambitious Subaru Prime Focus Spectrograph \citep[PFS,][]{2012SPIE.8446E..0YS}
survey with its powerful NIR capability 
promises to provide an unprecedented large high-redshift sample of
spectroscopic redshifts.

To fully exploit the greater statistical power
in future surveys, it is critical to control systematics.
In particular, the luminosity function slopes 
of source samples must be precisely known, 
in order to accurately 
extract the lensing signal from observations.
To do this, a homogeneous selection of background sources
is crucial.

To compute theoretical predictions, we have implicitly assumed that
mass is proportional to luminosity when constructing the above
composite cluster lensing mass profiles.
In details we must expect that the mass-to-light ratio varies as a
function of mass, redshift, and galaxy type.  In
\citet{2012A&A...542A...5C}, we fit the mass-to-light ratio as
function of mass and redshift for blue and red galaxies, using the
30-band COSMOS data \citep{2009ApJ...690.1236I}: in the mass bin
$10^{10.5}-10^{11.5} M_{\odot}$, we find the slope of the luminosity function 
changes from 0.44 to 0.69 depending of the galaxy type. This implies that, at most, the change
in luminosity keeping a constant mass would be about 0.1 magnitude
from redshift 0.4 to 0.7.  Given the parametrization of the luminosity
function employed here we can test how the slope of the number counts
would be affected by simply shifting the limiting magnitude by 0.1. In
practice, since we are concerned only with the relatively bright end
of the galaxy distribution, it is equivalent to a shift in $M_{\star}$
parameter. 

Furthermore, two additional sources of systematic errors could have
affected the determination of the source luminosity function by
\citet{2005A&A...439..863I}: the Poisson error and the cosmic
variance.  The former was estimated by \citet{2005A&A...439..863I},
and given as $\Delta M_{\star}\sim0.25$, which is correlated with the
faint-end slope $\alpha$, so that this estimate may be rather
pessimistic.  For the later, the effect of cosmic variance on the count
slope is more difficult to assess.  To obtain a crude estimate of this
effect we consider the value of $\Delta M_{\star}=0.02$, which was estimated by
\citet{2012MNRAS.420.1239L} for the GAMA survey 
from jackknife resampling of nine subregions each with 16\,deg$^2$.
Although GAMA is a lower-redshift survey compared to VVDS, 
the volume of each subregion is larger than the volume probed
over the 1~deg$^2$ field-of-view of VVDS at $z=0.5$, so that 
we have adopted a conservative estimate of 0.1.

Finally we add in quadrature all three sources of systematic
uncertainty, and reexamine our model predictions in the two extreme
cases, $M_{\star} \pm 0.29$.
In Fig.~\ref{fig:results_sys}, the resulting error region is indicated by the blue-shaded area
which is systematically less than the statistical error for the current dataset.

\section{Conclusions}

Using over $300\,000$ BOSS galaxies we have measured the mean redshift
of background galaxies behind large samples of SDSS galaxy
clusters. Our results show a net increase of the mean redshift behind
the clusters compared to that of the total sample, in line with
reasonable expectations for the effect of lens magnification. We have
tested four different cluster catalogs, the maxBCG, GMBCG, AMF and
WHL12 samples, with detection significance ranging from 2.8$\sigma$s
(GMBCG) to 4.9$\sigma$s (AMF), where the level of systematic errors is
expected to be negligible compared to our statistical errors. In order
to speed-up and ease the measurements of the mean redshift as function
of physical scale around clusters, we employed a sophisticated code
({\sc Swot}) that performs rapid parallel calculations and accurate
error estimations, making practical the handling of future samples
containing several millions of objects.

Based on precise measurements of luminosity functions from a number of
deep spectroscopic surveys, and on an accurate modelling of dark
matter halo profiles, we have compared our results to theoretical
predictions. We constructed the expected signal by summing the
individual contribution for each cluster given its richness from two
independently established mass-richness relationships corresponding to two
kinds of richness definition, and each calibrated with X-ray and weak
lensing data. For three samples out of four, the agreement with the
NFW profile is excellent.  The mean masses of the
clusters derived from our basic stacking analysis vary between the four 
large SDSS cluster samples in the range $1.18$--$1.81\time 10^{14}M_\odot$,
after allowing for the mass-richness relations appropriate for each
cluster sample and with a mean radial profile consistent with the observed
radial trend towards higher mean redshift at smaller cluster radius.
Only WHL12 showed a marginal deviation from the
expected level, suggesting an underestimated signal. Given the high density of
candidate clusters in this catalog compared to the others, this effect
may indicate a somewhat larger contamination of spurious clusters that
dilute the lensing signal in proportion to the fraction of false
detections.

Further investigation of possible sources of systematic error will be
feasible with a more thorough understanding of the source population
of the BOSS survey, in terms of colour and magnitude selection and
Eddington bias affecting the very bright end of the BOSS luminosity
function, arising from photometric error. With the current sample,
given our relatively large statistical errors we conclude that the
systematic errors in the measurements should be negligible at present
but that with the completion of the BOSS survey we will reach an
increased level of precision that may warrant closer scrutiny of the
details of the source selection function and possible redshift
uncertainties for the foreground cluster populations. We believe that
the method is very robust against systematic errors, even when we
increase the statistical power in future experiments such as bigBOSS,
PFS and EUCLID (which will observe background sources 25 times as
dense and over 5 times the area of this study) with millions of
spectroscopic/grism redshifts, at which point this method will be
powerful in its own right for defining cluster mass-concentration
relations and the evolution of the cluster mass function. These
results may be compared with independent estimates of magnification
bias from faint number counts and with cosmic shear measurements for
which the sources of systematic error are very different.

\acknowledgments
We acknowledge Yen-Ting Lin for fruitful discussions and comments, and for 
providing us with the calibration formula for MCXC masses.

Funding for SDSS-III has been provided by the Alfred P. Sloan Foundation, the Participating Institutions, the National Science Foundation, and the U.S. Department of Energy Office of Science. The SDSS-III web site is \url{http://www.sdss3.org/}.

SDSS-III is managed by the Astrophysical Research Consortium for the Participating Institutions of the SDSS-III Collaboration including the University of Arizona, the Brazilian Participation Group, Brookhaven National Laboratory, University of Cambridge, Carnegie Mellon University, University of Florida, the French Participation Group, the German Participation Group, Harvard University, the Instituto de Astrofisica de Canarias, the Michigan State/Notre Dame/JINA Participation Group, Johns Hopkins University, Lawrence Berkeley National Laboratory, Max Planck Institute for Astrophysics, Max Planck Institute for Extraterrestrial Physics, New Mexico State University, New York University, Ohio State University, Pennsylvania State University, University of Portsmouth, Princeton University, the Spanish Participation Group, University of Tokyo, University of Utah, Vanderbilt University, University of Virginia, University of Washington, and Yale University.

The work is partially supported by the National Science Council of Taiwan
under the grant NSC97-2112-M-001-020-MY3 and by
the Academia Sinica Career Development Award.

\bibliographystyle{apj}
\bibliographystyle{aa}
\bibliography{references}

\begin{thebibliography}{67}
\expandafter\ifx\csname natexlab\endcsname\relax\def\natexlab#1{#1}\fi

\bibitem[{{Bartelmann} \& {Schneider}(2001)}]{2001PhR...340..291B}
{Bartelmann}, M. \& {Schneider}, P. 2001, \physrep, 340, 291

\bibitem[{{Benjamin} {et~al.}(2012){Benjamin}, {Van Waerbeke}, {Heymans},
  {Kilbinger}, {Erben}, {Hildebrandt}, {Hoekstra}, {Kitching}, {Mellier},
  {Miller}, {Rowe}, {Schrabback}, {Simpson}, {Coupon}, {Fu},
  {Harnois-D{\'e}raps}, {Hudson}, {Kuijken}, {Semboloni}, {Vafaei}, \&
  {Velander}}]{2012arXiv1212.3327B}
{Benjamin}, J., {Van Waerbeke}, L., {Heymans}, C., {et~al.} 2012, ArXiv
  e-prints

\bibitem[{{Bhattacharya} {et~al.}(2013){Bhattacharya}, {Habib}, {Heitmann}, \&
  {Vikhlinin}}]{2013ApJ...766...32B}
{Bhattacharya}, S., {Habib}, S., {Heitmann}, K., \& {Vikhlinin}, A. 2013, \apj,
  766, 32

\bibitem[{{Broadhurst} {et~al.}(2005){Broadhurst}, {Takada}, {Umetsu}, {Kong},
  {Arimoto}, {Chiba}, \& {Futamase}}]{2005ApJ...619L.143B}
{Broadhurst}, T., {Takada}, M., {Umetsu}, K., {et~al.} 2005, \apjl, 619, L143

\bibitem[{{Broadhurst} {et~al.}(2008){Broadhurst}, {Umetsu}, {Medezinski},
  {Oguri}, \& {Rephaeli}}]{2008ApJ...685L...9B}
{Broadhurst}, T., {Umetsu}, K., {Medezinski}, E., {Oguri}, M., \& {Rephaeli},
  Y. 2008, \apjl, 685, L9

\bibitem[{{Broadhurst} {et~al.}(1995){Broadhurst}, {Taylor}, \&
  {Peacock}}]{1995ApJ...438...49B}
{Broadhurst}, T.~J., {Taylor}, A.~N., \& {Peacock}, J.~A. 1995, \apj, 438, 49

\bibitem[{{Coe} {et~al.}(2012){Coe}, {Umetsu}, {Zitrin}, {Donahue},
  {Medezinski}, {Postman}, {Carrasco}, {Anguita}, {Geller}, {Rines},
  {Diaferio}, {Kurtz}, {Bradley}, {Koekemoer}, {Zheng}, {Nonino}, {Molino},
  {Mahdavi}, {Lemze}, {Infante}, {Ogaz}, {Melchior}, {Host}, {Ford}, {Grillo},
  {Rosati}, {Jim{\'e}nez-Teja}, {Moustakas}, {Broadhurst}, {Ascaso}, {Lahav},
  {Bartelmann}, {Ben{\'{\i}}tez}, {Bouwens}, {Graur}, {Graves}, {Jha},
  {Jouvel}, {Kelson}, {Moustakas}, {Maoz}, {Meneghetti}, {Merten}, {Riess},
  {Rodney}, \& {Seitz}}]{2012ApJ...757...22C}
{Coe}, D., {Umetsu}, K., {Zitrin}, A., {et~al.} 2012, \apj, 757, 22

\bibitem[{{Coupon} {et~al.}(2012){Coupon}, {Kilbinger}, {McCracken}, {Ilbert},
  {Arnouts}, {Mellier}, {Abbas}, {de la Torre}, {Goranova}, {Hudelot}, {Kneib},
  \& {Le F{\`e}vre}}]{2012A&A...542A...5C}
{Coupon}, J., {Kilbinger}, M., {McCracken}, H.~J., {et~al.} 2012, \aap, 542, A5

\bibitem[{{Davis} \& {Peebles}(1983)}]{1983ApJ...267..465D}
{Davis}, M. \& {Peebles}, P.~J.~E. 1983, \apj, 267, 465

\bibitem[{{Dawson} {et~al.}(2013){Dawson}, {Schlegel}, {Ahn}, {Anderson},
  {Aubourg}, {Bailey}, {Barkhouser}, {Bautista}, {Beifiori}, {Berlind},
  {Bhardwaj}, {Bizyaev}, {Blake}, {Blanton}, {Blomqvist}, {Bolton}, {Borde},
  {Bovy}, {Brandt}, {Brewington}, {Brinkmann}, {Brown}, {Brownstein}, {Bundy},
  {Busca}, {Carithers}, {Carnero}, {Carr}, {Chen}, {Comparat}, {Connolly},
  {Cope}, {Croft}, {Cuesta}, {da Costa}, {Davenport}, {Delubac}, {de Putter},
  {Dhital}, {Ealet}, {Ebelke}, {Eisenstein}, {Escoffier}, {Fan}, {Filiz Ak},
  {Finley}, {Font-Ribera}, {G{\'e}nova-Santos}, {Gunn}, {Guo}, {Haggard},
  {Hall}, {Hamilton}, {Harris}, {Harris}, {Ho}, {Hogg}, {Holder}, {Honscheid},
  {Huehnerhoff}, {Jordan}, {Jordan}, {Kauffmann}, {Kazin}, {Kirkby}, {Klaene},
  {Kneib}, {Le Goff}, {Lee}, {Long}, {Loomis}, {Lundgren}, {Lupton}, {Maia},
  {Makler}, {Malanushenko}, {Malanushenko}, {Mandelbaum}, {Manera}, {Maraston},
  {Margala}, {Masters}, {McBride}, {McDonald}, {McGreer}, {McMahon}, {Mena},
  {Miralda-Escud{\'e}}, {Montero-Dorta}, {Montesano}, {Muna}, {Myers},
  {Naugle}, {Nichol}, {Noterdaeme}, {Nuza}, {Olmstead}, {Oravetz}, {Oravetz},
  {Owen}, {Padmanabhan}, {Palanque-Delabrouille}, {Pan}, {Parejko},
  {P{\^a}ris}, {Percival}, {P{\'e}rez-Fournon}, {P{\'e}rez-R{\`a}fols},
  {Petitjean}, {Pfaffenberger}, {Pforr}, {Pieri}, {Prada}, {Price-Whelan},
  {Raddick}, {Rebolo}, {Rich}, {Richards}, {Rockosi}, {Roe}, {Ross}, {Ross},
  {Rossi}, {Rubi{\~n}o-Martin}, {Samushia}, {S{\'a}nchez}, {Sayres}, {Schmidt},
  {Schneider}, {Sc{\'o}ccola}, {Seo}, {Shelden}, {Sheldon}, {Shen}, {Shu},
  {Slosar}, {Smee}, {Snedden}, {Stauffer}, {Steele}, {Strauss}, {Streblyanska},
  {Suzuki}, {Swanson}, {Tal}, {Tanaka}, {Thomas}, {Tinker}, {Tojeiro},
  {Tremonti}, {Vargas Maga{\~n}a}, {Verde}, {Viel}, {Wake}, {Watson}, {Weaver},
  {Weinberg}, {Weiner}, {West}, {White}, {Wood-Vasey}, {Yeche}, {Zehavi},
  {Zhao}, \& {Zheng}}]{2013AJ....145...10D}
{Dawson}, K.~S., {Schlegel}, D.~J., {Ahn}, C.~P., {et~al.} 2013, \aj, 145, 10

\bibitem[{{Dong} {et~al.}(2008){Dong}, {Pierpaoli}, {Gunn}, \&
  {Wechsler}}]{2008ApJ...676..868D}
{Dong}, F., {Pierpaoli}, E., {Gunn}, J.~E., \& {Wechsler}, R.~H. 2008, \apj,
  676, 868

\bibitem[{{Faber} {et~al.}(2007){Faber}, {Willmer}, {Wolf}, {Koo}, {Weiner},
  {Newman}, {Im}, {Coil}, {Conroy}, {Cooper}, {Davis}, {Finkbeiner}, {Gerke},
  {Gebhardt}, {Groth}, {Guhathakurta}, {Harker}, {Kaiser}, {Kassin},
  {Kleinheinrich}, {Konidaris}, {Kron}, {Lin}, {Luppino}, {Madgwick},
  {Meisenheimer}, {Noeske}, {Phillips}, {Sarajedini}, {Schiavon}, {Simard},
  {Szalay}, {Vogt}, \& {Yan}}]{2007ApJ...665..265F}
{Faber}, S.~M., {Willmer}, C.~N.~A., {Wolf}, C., {et~al.} 2007, \apj, 665, 265

\bibitem[{{Ford} {et~al.}(2012){Ford}, {Hildebrandt}, {Van Waerbeke},
  {Leauthaud}, {Capak}, {Finoguenov}, {Tanaka}, {George}, \&
  {Rhodes}}]{2012ApJ...754..143F}
{Ford}, J., {Hildebrandt}, H., {Van Waerbeke}, L., {et~al.} 2012, \apj, 754,
  143

\bibitem[{{Fukugita} {et~al.}(1996){Fukugita}, {Ichikawa}, {Gunn}, {Doi},
  {Shimasaku}, \& {Schneider}}]{1996AJ....111.1748F}
{Fukugita}, M., {Ichikawa}, T., {Gunn}, J.~E., {et~al.} 1996, \aj, 111, 1748

\bibitem[{{George} {et~al.}(2012){George}, {Leauthaud}, {Bundy}, {Finoguenov},
  {Ma}, {Rykoff}, {Tinker}, {Wechsler}, {Massey}, \&
  {Mei}}]{2012ApJ...757....2G}
{George}, M.~R., {Leauthaud}, A., {Bundy}, K., {et~al.} 2012, \apj, 757, 2

\bibitem[{{Gunn} {et~al.}(1998){Gunn}, {Carr}, {Rockosi}, {Sekiguchi}, {Berry},
  {Elms}, {de Haas}, {Ivezi{\'c}}, {Knapp}, {Lupton}, {Pauls}, {Simcoe},
  {Hirsch}, {Sanford}, {Wang}, {York}, {Harris}, {Annis}, {Bartozek},
  {Boroski}, {Bakken}, {Haldeman}, {Kent}, {Holm}, {Holmgren}, {Petravick},
  {Prosapio}, {Rechenmacher}, {Doi}, {Fukugita}, {Shimasaku}, {Okada}, {Hull},
  {Siegmund}, {Mannery}, {Blouke}, {Heidtman}, {Schneider}, {Lucinio}, \&
  {Brinkman}}]{1998AJ....116.3040G}
{Gunn}, J.~E., {Carr}, M., {Rockosi}, C., {et~al.} 1998, \aj, 116, 3040

\bibitem[{{Hao} {et~al.}(2010){Hao}, {McKay}, {Koester}, {Rykoff}, {Rozo},
  {Annis}, {Wechsler}, {Evrard}, {Siegel}, {Becker}, {Busha}, {Gerdes},
  {Johnston}, \& {Sheldon}}]{2010ApJS..191..254H}
{Hao}, J., {McKay}, T.~A., {Koester}, B.~P., {et~al.} 2010, \apjs, 191, 254

\bibitem[{{Heymans} {et~al.}(2012){Heymans}, {Van Waerbeke}, {Miller}, {Erben},
  {Hildebrandt}, {Hoekstra}, {Kitching}, {Mellier}, {Simon}, {Bonnett},
  {Coupon}, {Fu}, {Harnois D{\'e}raps}, {Hudson}, {Kilbinger}, {Kuijken},
  {Rowe}, {Schrabback}, {Semboloni}, {van Uitert}, {Vafaei}, \&
  {Velander}}]{2012MNRAS.427..146H}
{Heymans}, C., {Van Waerbeke}, L., {Miller}, L., {et~al.} 2012, \mnras, 427,
  146

\bibitem[{{Hildebrandt} {et~al.}(2011){Hildebrandt}, {Muzzin}, {Erben},
  {Hoekstra}, {Kuijken}, {Surace}, {van Waerbeke}, {Wilson}, \&
  {Yee}}]{2011ApJ...733L..30H}
{Hildebrandt}, H., {Muzzin}, A., {Erben}, T., {et~al.} 2011, \apjl, 733, L30

\bibitem[{{Hildebrandt} {et~al.}(2013){Hildebrandt}, {van Waerbeke}, {Scott},
  {B{\'e}thermin}, {Bock}, {Clements}, {Conley}, {Cooray}, {Dunlop}, {Eales},
  {Erben}, {Farrah}, {Franceschini}, {Glenn}, {Halpern}, {Heinis}, {Ivison},
  {Marsden}, {Oliver}, {Page}, {P{\'e}rez-Fournon}, {Smith}, {Rowan-Robinson},
  {Valtchanov}, {van der Burg}, {Vieira}, {Viero}, \&
  {Wang}}]{2013MNRAS.429.3230H}
{Hildebrandt}, H., {van Waerbeke}, L., {Scott}, D., {et~al.} 2013, \mnras, 429,
  3230

\bibitem[{{Hinshaw} {et~al.}(2009){Hinshaw}, {Weiland}, {Hill}, {Odegard},
  {Larson}, {Bennett}, {Dunkley}, {Gold}, {Greason}, {Jarosik}, {Komatsu},
  {Nolta}, {Page}, {Spergel}, {Wollack}, {Halpern}, {Kogut}, {Limon}, {Meyer},
  {Tucker}, \& {Wright}}]{2009ApJS..180..225H}
{Hinshaw}, G., {Weiland}, J.~L., {Hill}, R.~S., {et~al.} 2009, \apjs, 180, 225

\bibitem[{{Hu} \& {Kravtsov}(2003)}]{2003ApJ...584..702H}
{Hu}, W. \& {Kravtsov}, A.~V. 2003, \apj, 584, 702

\bibitem[{{Ilbert} {et~al.}(2009){Ilbert}, {Capak}, {Salvato}, {Aussel},
  {McCracken}, {Sanders}, {Scoville}, {Kartaltepe}, {Arnouts}, {Le Floc'h},
  {Mobasher}, {Taniguchi}, {Lamareille}, {Leauthaud}, {Sasaki}, {Thompson},
  {Zamojski}, {Zamorani}, {Bardelli}, {Bolzonella}, {Bongiorno}, {Brusa},
  {Caputi}, {Carollo}, {Contini}, {Cook}, {Coppa}, {Cucciati}, {de la Torre},
  {de Ravel}, {Franzetti}, {Garilli}, {Hasinger}, {Iovino}, {Kampczyk},
  {Kneib}, {Knobel}, {Kovac}, {Le Borgne}, {Le Brun}, {F{\`e}vre}, {Lilly},
  {Looper}, {Maier}, {Mainieri}, {Mellier}, {Mignoli}, {Murayama}, {Pell{\`o}},
  {Peng}, {P{\'e}rez-Montero}, {Renzini}, {Ricciardelli}, {Schiminovich},
  {Scodeggio}, {Shioya}, {Silverman}, {Surace}, {Tanaka}, {Tasca}, {Tresse},
  {Vergani}, \& {Zucca}}]{2009ApJ...690.1236I}
{Ilbert}, O., {Capak}, P., {Salvato}, M., {et~al.} 2009, \apj, 690, 1236

\bibitem[{{Ilbert} {et~al.}(2005){Ilbert}, {Tresse}, {Zucca}, {Bardelli},
  {Arnouts}, {Zamorani}, {Pozzetti}, {Bottini}, {Garilli}, {Le Brun}, {Le
  F{\`e}vre}, {Maccagni}, {Picat}, {Scaramella}, {Scodeggio}, {Vettolani},
  {Zanichelli}, {Adami}, {Arnaboldi}, {Bolzonella}, {Cappi}, {Charlot},
  {Contini}, {Foucaud}, {Franzetti}, {Gavignaud}, {Guzzo}, {Iovino},
  {McCracken}, {Marano}, {Marinoni}, {Mathez}, {Mazure}, {Meneux}, {Merighi},
  {Paltani}, {Pello}, {Pollo}, {Radovich}, {Bondi}, {Bongiorno}, {Busarello},
  {Ciliegi}, {Lamareille}, {Mellier}, {Merluzzi}, {Ripepi}, \&
  {Rizzo}}]{2005A&A...439..863I}
{Ilbert}, O., {Tresse}, L., {Zucca}, E., {et~al.} 2005, \aap, 439, 863

\bibitem[{{Johnston} {et~al.}(2007){Johnston}, {Sheldon}, {Tasitsiomi},
  {Frieman}, {Wechsler}, \& {McKay}}]{2007ApJ...656...27J}
{Johnston}, D.~E., {Sheldon}, E.~S., {Tasitsiomi}, A., {et~al.} 2007, \apj,
  656, 27

\bibitem[{{Kaiser} {et~al.}(1995){Kaiser}, {Squires}, \&
  {Broadhurst}}]{1995ApJ...449..460K}
{Kaiser}, N., {Squires}, G., \& {Broadhurst}, T. 1995, \apj, 449, 460

\bibitem[{{Kilbinger} {et~al.}(2013){Kilbinger}, {Fu}, {Heymans}, {Simpson},
  {Benjamin}, {Erben}, {Harnois-D{\'e}raps}, {Hoekstra}, {Hildebrandt},
  {Kitching}, {Mellier}, {Miller}, {Van Waerbeke}, {Benabed}, {Bonnett},
  {Coupon}, {Hudson}, {Kuijken}, {Rowe}, {Schrabback}, {Semboloni}, {Vafaei},
  \& {Velander}}]{2013MNRAS.tmp..735K}
{Kilbinger}, M., {Fu}, L., {Heymans}, C., {et~al.} 2013, \mnras

\bibitem[{{Koester} {et~al.}(2007){Koester}, {McKay}, {Annis}, {Wechsler},
  {Evrard}, {Bleem}, {Becker}, {Johnston}, {Sheldon}, {Nichol}, {Miller},
  {Scranton}, {Bahcall}, {Barentine}, {Brewington}, {Brinkmann}, {Harvanek},
  {Kleinman}, {Krzesinski}, {Long}, {Nitta}, {Schneider}, {Sneddin}, {Voges},
  \& {York}}]{2007ApJ...660..239K}
{Koester}, B.~P., {McKay}, T.~A., {Annis}, J., {et~al.} 2007, \apj, 660, 239

\bibitem[{{Laureijs} {et~al.}(2011){Laureijs}, {Amiaux}, {Arduini},
  {Augu{\`e}res}, {Brinchmann}, {Cole}, {Cropper}, {Dabin}, {Duvet}, {Ealet},
  \& et~al.}]{2011arXiv1110.3193L}
{Laureijs}, R., {Amiaux}, J., {Arduini}, S., {et~al.} 2011, ArXiv e-prints

\bibitem[{{Le F{\`e}vre} {et~al.}(2005){Le F{\`e}vre}, {Vettolani}, {Garilli},
  {Tresse}, {Bottini}, {Le Brun}, {Maccagni}, {Picat}, {Scaramella},
  {Scodeggio}, {Zanichelli}, {Adami}, {Arnaboldi}, {Arnouts}, {Bardelli},
  {Bolzonella}, {Cappi}, {Charlot}, {Ciliegi}, {Contini}, {Foucaud},
  {Franzetti}, {Gavignaud}, {Guzzo}, {Ilbert}, {Iovino}, {McCracken}, {Marano},
  {Marinoni}, {Mathez}, {Mazure}, {Meneux}, {Merighi}, {Paltani}, {Pell{\`o}},
  {Pollo}, {Pozzetti}, {Radovich}, {Zamorani}, {Zucca}, {Bondi}, {Bongiorno},
  {Busarello}, {Lamareille}, {Mellier}, {Merluzzi}, {Ripepi}, \&
  {Rizzo}}]{2005A&A...439..845L}
{Le F{\`e}vre}, O., {Vettolani}, G., {Garilli}, B., {et~al.} 2005, \aap, 439,
  845

\bibitem[{{Leauthaud} {et~al.}(2010){Leauthaud}, {Finoguenov}, {Kneib},
  {Taylor}, {Massey}, {Rhodes}, {Ilbert}, {Bundy}, {Tinker}, {George}, {Capak},
  {Koekemoer}, {Johnston}, {Zhang}, {Cappelluti}, {Ellis}, {Elvis}, {Giodini},
  {Heymans}, {Le F{\`e}vre}, {Lilly}, {McCracken}, {Mellier},
  {R{\'e}fr{\'e}gier}, {Salvato}, {Scoville}, {Smoot}, {Tanaka}, {Van
  Waerbeke}, \& {Wolk}}]{2010ApJ...709...97L}
{Leauthaud}, A., {Finoguenov}, A., {Kneib}, J.-P., {et~al.} 2010, \apj, 709, 97

\bibitem[{{Loveday} {et~al.}(2012){Loveday}, {Norberg}, {Baldry}, {Driver},
  {Hopkins}, {Peacock}, {Bamford}, {Liske}, {Bland-Hawthorn}, {Brough},
  {Brown}, {Cameron}, {Conselice}, {Croom}, {Frenk}, {Gunawardhana}, {Hill},
  {Jones}, {Kelvin}, {Kuijken}, {Nichol}, {Parkinson}, {Phillipps}, {Pimbblet},
  {Popescu}, {Prescott}, {Robotham}, {Sharp}, {Sutherland}, {Taylor}, {Thomas},
  {Tuffs}, {van Kampen}, \& {Wijesinghe}}]{2012MNRAS.420.1239L}
{Loveday}, J., {Norberg}, P., {Baldry}, I.~K., {et~al.} 2012, \mnras, 420, 1239

\bibitem[{{Mandelbaum} {et~al.}(2008){Mandelbaum}, {Seljak}, \&
  {Hirata}}]{2008JCAP...08..006M}
{Mandelbaum}, R., {Seljak}, U., \& {Hirata}, C.~M. 2008, jcap, 8, 6

\bibitem[{{Maraston} {et~al.}(2012){Maraston}, {Pforr}, {Henriques}, {Thomas},
  {Wake}, {Brownstein}, {Capozzi}, {Bundy}, {Skibba}, {Beifiori}, {Nichol},
  {Edmondson}, {Schneider}, {Chen}, {Masters}, {Steele}, {Bolton}, {York},
  {Bizyaev}, {Brewington}, {Malanushenko}, {Malanushenko}, {Snedden},
  {Oravetz}, {Pan}, {Shelden}, \& {Simmons}}]{2012arXiv1207.6114M}
{Maraston}, C., {Pforr}, J., {Henriques}, B.~M., {et~al.} 2012, ArXiv e-prints

\bibitem[{{Marriage} {et~al.}(2011){Marriage}, {Acquaviva}, {Ade}, {Aguirre},
  {Amiri}, {Appel}, {Barrientos}, {Battistelli}, {Bond}, {Brown}, {Burger},
  {Chervenak}, {Das}, {Devlin}, {Dicker}, {Bertrand Doriese}, {Dunkley},
  {D{\"u}nner}, {Essinger-Hileman}, {Fisher}, {Fowler}, {Hajian}, {Halpern},
  {Hasselfield}, {Hern{\'a}ndez-Monteagudo}, {Hilton}, {Hilton}, {Hincks},
  {Hlozek}, {Huffenberger}, {Handel Hughes}, {Hughes}, {Infante}, {Irwin},
  {Baptiste Juin}, {Kaul}, {Klein}, {Kosowsky}, {Lau}, {Limon}, {Lin},
  {Lupton}, {Marsden}, {Martocci}, {Mauskopf}, {Menanteau}, {Moodley},
  {Moseley}, {Netterfield}, {Niemack}, {Nolta}, {Page}, {Parker}, {Partridge},
  {Quintana}, {Reese}, {Reid}, {Sehgal}, {Sherwin}, {Sievers}, {Spergel},
  {Staggs}, {Swetz}, {Switzer}, {Thornton}, {Trac}, {Tucker}, {Warne},
  {Wilson}, {Wollack}, \& {Zhao}}]{2011ApJ...737...61M}
{Marriage}, T.~A., {Acquaviva}, V., {Ade}, P.~A.~R., {et~al.} 2011, \apj, 737,
  61

\bibitem[{{Massey} {et~al.}(2007){Massey}, {Rhodes}, {Leauthaud}, {Capak},
  {Ellis}, {Koekemoer}, {R{\'e}fr{\'e}gier}, {Scoville}, {Taylor}, {Albert},
  {Berg{\'e}}, {Heymans}, {Johnston}, {Kneib}, {Mellier}, {Mobasher},
  {Semboloni}, {Shopbell}, {Tasca}, \& {Van Waerbeke}}]{2007ApJS..172..239M}
{Massey}, R., {Rhodes}, J., {Leauthaud}, A., {et~al.} 2007, \apjs, 172, 239

\bibitem[{{M{\'e}nard} {et~al.}(2010){M{\'e}nard}, {Scranton}, {Fukugita}, \&
  {Richards}}]{2010MNRAS.405.1025M}
{M{\'e}nard}, B., {Scranton}, R., {Fukugita}, M., \& {Richards}, G. 2010,
  \mnras, 405, 1025

\bibitem[{{Miller} {et~al.}(2013){Miller}, {Heymans}, {Kitching}, {van
  Waerbeke}, {Erben}, {Hildebrandt}, {Hoekstra}, {Mellier}, {Rowe}, {Coupon},
  {Dietrich}, {Fu}, {Harnois-D{\'e}raps}, {Hudson}, {Kilbinger}, {Kuijken},
  {Schrabback}, {Semboloni}, {Vafaei}, \& {Velander}}]{2013MNRAS.429.2858M}
{Miller}, L., {Heymans}, C., {Kitching}, T.~D., {et~al.} 2013, \mnras, 429,
  2858

\bibitem[{{Miyazaki} {et~al.}(2012){Miyazaki}, {Komiyama}, {Nakaya}, {Kamata},
  {Doi}, {Hamana}, {Karoji}, {Furusawa}, {Kawanomoto}, {Morokuma}, {Ishizuka},
  {Nariai}, {Tanaka}, {Uraguchi}, {Utsumi}, {Obuchi}, {Okura}, {Oguri},
  {Takata}, {Tomono}, {Kurakami}, {Namikawa}, {Usuda}, {Yamanoi}, {Terai},
  {Uekiyo}, {Yamada}, {Koike}, {Aihara}, {Fujimori}, {Mineo}, {Miyatake},
  {Yasuda}, {Nishizawa}, {Saito}, {Tanaka}, {Uchida}, {Katayama}, {Wang},
  {Chen}, {Lupton}, {Loomis}, {Bickerton}, {Price}, {Gunn}, {Suzuki},
  {Miyazaki}, {Muramatsu}, {Yamamoto}, {Endo}, {Ezaki}, {Itoh}, {Miwa},
  {Yokota}, {Matsuda}, {Ebinuma}, \& {Takeshi}}]{2012SPIE.8446E..0ZM}
{Miyazaki}, S., {Komiyama}, Y., {Nakaya}, H., {et~al.} 2012, in Society of
  Photo-Optical Instrumentation Engineers (SPIE) Conference Series, Vol. 8446,
  Society of Photo-Optical Instrumentation Engineers (SPIE) Conference Series

\bibitem[{{More} {et~al.}(2011){More}, {van den Bosch}, {Cacciato}, {Skibba},
  {Mo}, \& {Yang}}]{2011MNRAS.410..210M}
{More}, S., {van den Bosch}, F.~C., {Cacciato}, M., {et~al.} 2011, \mnras, 410,
  210

\bibitem[{{Navarro} {et~al.}(1997){Navarro}, {Frenk}, \&
  {White}}]{1997ApJ...490..493N}
{Navarro}, J.~F., {Frenk}, C.~S., \& {White}, S.~D.~M. 1997, \apj, 490, 493

\bibitem[{{Oguri} \& {Hamana}(2011)}]{2011MNRAS.414.1851O}
{Oguri}, M. \& {Hamana}, T. 2011, \mnras, 414, 1851

\bibitem[{{Okabe} {et~al.}(2013){Okabe}, {Smith}, {Umetsu}, {Takada}, \&
  {Futamase}}]{2013arXiv1302.2728O}
{Okabe}, N., {Smith}, G.~P., {Umetsu}, K., {Takada}, M., \& {Futamase}, T.
  2013, ArXiv e-prints

\bibitem[{{Okabe} {et~al.}(2010){Okabe}, {Takada}, {Umetsu}, {Futamase}, \&
  {Smith}}]{2010PASJ...62..811O}
{Okabe}, N., {Takada}, M., {Umetsu}, K., {Futamase}, T., \& {Smith}, G.~P.
  2010, \pasj, 62, 811

\bibitem[{{Oyaizu} {et~al.}(2008){Oyaizu}, {Lima}, {Cunha}, {Lin}, {Frieman},
  \& {Sheldon}}]{2008ApJ...674..768O}
{Oyaizu}, H., {Lima}, M., {Cunha}, C.~E., {et~al.} 2008, \apj, 674, 768

\bibitem[{{Piffaretti} {et~al.}(2011){Piffaretti}, {Arnaud}, {Pratt},
  {Pointecouteau}, \& {Melin}}]{2011A&A...534A.109P}
{Piffaretti}, R., {Arnaud}, M., {Pratt}, G.~W., {Pointecouteau}, E., \&
  {Melin}, J.-B. 2011, \aap, 534, A109

\bibitem[{{Rozo} {et~al.}(2009){Rozo}, {Rykoff}, {Evrard}, {Becker}, {McKay},
  {Wechsler}, {Koester}, {Hao}, {Hansen}, {Sheldon}, {Johnston}, {Annis}, \&
  {Frieman}}]{2009ApJ...699..768R}
{Rozo}, E., {Rykoff}, E.~S., {Evrard}, A., {et~al.} 2009, \apj, 699, 768

\bibitem[{{Rozo} {et~al.}(2010){Rozo}, {Wechsler}, {Rykoff}, {Annis}, {Becker},
  {Evrard}, {Frieman}, {Hansen}, {Hao}, {Johnston}, {Koester}, {McKay},
  {Sheldon}, \& {Weinberg}}]{2010ApJ...708..645R}
{Rozo}, E., {Wechsler}, R.~H., {Rykoff}, E.~S., {et~al.} 2010, \apj, 708, 645

\bibitem[{{Rykoff} {et~al.}(2008){Rykoff}, {McKay}, {Becker}, {Evrard},
  {Johnston}, {Koester}, {Rozo}, {Sheldon}, \&
  {Wechsler}}]{2008ApJ...675.1106R}
{Rykoff}, E.~S., {McKay}, T.~A., {Becker}, M.~R., {et~al.} 2008, \apj, 675,
  1106

\bibitem[{{Schechter}(1976)}]{1976ApJ...203..297S}
{Schechter}, P. 1976, \apj, 203, 297

\bibitem[{{Schlegel} {et~al.}(2009){Schlegel}, {White}, \&
  {Eisenstein}}]{2009astro2010S.314S}
{Schlegel}, D., {White}, M., \& {Eisenstein}, D. 2009, in Astronomy, Vol. 2010,
  astro2010: The Astronomy and Astrophysics Decadal Survey, 314

\bibitem[{{Sheldon} {et~al.}(2009){Sheldon}, {Johnston}, {Scranton}, {Koester},
  {McKay}, {Oyaizu}, {Cunha}, {Lima}, {Lin}, {Frieman}, {Wechsler}, {Annis},
  {Mandelbaum}, {Bahcall}, \& {Fukugita}}]{2009ApJ...703.2217S}
{Sheldon}, E.~S., {Johnston}, D.~E., {Scranton}, R., {et~al.} 2009, \apj, 703,
  2217

\bibitem[{{Sugai} {et~al.}(2012){Sugai}, {Karoji}, {Takato}, {Tamura},
  {Shimono}, {Ohyama}, {Ueda}, {Ling}, {Vital de Arruda}, {Barkhouser},
  {Bennett}, {Bickerton}, {Braun}, {Bruno}, {Carr}, {Batista de Carvalho
  Oliveira}, {Chang}, {Chen}, {Dekany}, {Pereira Dominici}, {Ellis}, {Fisher},
  {Gunn}, {Heckman}, {Ho}, {Hu}, {Jaquet}, {Karr}, {Kimura}, {Le F{\`e}vre},
  {Le Mignant}, {Loomis}, {Lupton}, {Madec}, {Marrara}, {Martin}, {Murayama},
  {Cesar de Oliveira}, {Mendes de Oliveira}, {Souza de Oliveira}, {Orndorff},
  {de Paiva Vila{\c c}a}, {Macanhan}, {Prieto}, {Bispo dos Santos}, {Seiffert},
  {Smee}, {Smith}, {Sodr{\'e}}, {Spergel}, {Surace}, {Vives}, {Wang}, \&
  {Yan}}]{2012SPIE.8446E..0YS}
{Sugai}, H., {Karoji}, H., {Takato}, N., {et~al.} 2012, in Society of
  Photo-Optical Instrumentation Engineers (SPIE) Conference Series, Vol. 8446,
  Society of Photo-Optical Instrumentation Engineers (SPIE) Conference Series

\bibitem[{{Szabo} {et~al.}(2011){Szabo}, {Pierpaoli}, {Dong}, {Pipino}, \&
  {Gunn}}]{2011ApJ...736...21S}
{Szabo}, T., {Pierpaoli}, E., {Dong}, F., {Pipino}, A., \& {Gunn}, J. 2011,
  \apj, 736, 21

\bibitem[{{Taylor} {et~al.}(2004){Taylor}, {Bacon}, {Gray}, {Wolf},
  {Meisenheimer}, {Dye}, {Borch}, {Kleinheinrich}, {Kovacs}, \&
  {Wisotzki}}]{2004MNRAS.353.1176T}
{Taylor}, A.~N., {Bacon}, D.~J., {Gray}, M.~E., {et~al.} 2004, \mnras, 353,
  1176

\bibitem[{{Umetsu} \& {Broadhurst}(2008)}]{2008ApJ...684..177U}
{Umetsu}, K. \& {Broadhurst}, T. 2008, \apj, 684, 177

\bibitem[{{Umetsu} {et~al.}(2011{\natexlab{a}}){Umetsu}, {Broadhurst},
  {Zitrin}, {Medezinski}, {Coe}, \& {Postman}}]{2011ApJ...738...41U}
{Umetsu}, K., {Broadhurst}, T., {Zitrin}, A., {et~al.} 2011{\natexlab{a}},
  \apj, 738, 41

\bibitem[{{Umetsu} {et~al.}(2011{\natexlab{b}}){Umetsu}, {Broadhurst},
  {Zitrin}, {Medezinski}, \& {Hsu}}]{2011ApJ...729..127U}
{Umetsu}, K., {Broadhurst}, T., {Zitrin}, A., {Medezinski}, E., \& {Hsu}, L.-Y.
  2011{\natexlab{b}}, \apj, 729, 127

\bibitem[{{Umetsu} {et~al.}(2012){Umetsu}, {Medezinski}, {Nonino}, {Merten},
  {Zitrin}, {Molino}, {Grillo}, {Carrasco}, {Donahue}, {Mahdavi}, {Coe},
  {Postman}, {Koekemoer}, {Czakon}, {Sayers}, {Mroczkowski}, {Golwala}, {Koch},
  {Lin}, {Molnar}, {Rosati}, {Balestra}, {Mercurio}, {Scodeggio}, {Biviano},
  {Anguita}, {Infante}, {Seidel}, {Sendra}, {Jouvel}, {Host}, {Lemze},
  {Broadhurst}, {Meneghetti}, {Moustakas}, {Bartelmann}, {Ben{\'{\i}}tez},
  {Bouwens}, {Bradley}, {Ford}, {Jim{\'e}nez-Teja}, {Kelson}, {Lahav},
  {Melchior}, {Moustakas}, {Ogaz}, {Seitz}, \& {Zheng}}]{2012ApJ...755...56U}
{Umetsu}, K., {Medezinski}, E., {Nonino}, M., {et~al.} 2012, \apj, 755, 56

\bibitem[{{Van Waerbeke} {et~al.}(2010){Van Waerbeke}, {Hildebrandt}, {Ford},
  \& {Milkeraitis}}]{2010ApJ...723L..13V}
{Van Waerbeke}, L., {Hildebrandt}, H., {Ford}, J., \& {Milkeraitis}, M. 2010,
  \apjl, 723, L13

\bibitem[{{Vikhlinin} {et~al.}(2009){Vikhlinin}, {Burenin}, {Ebeling},
  {Forman}, {Hornstrup}, {Jones}, {Kravtsov}, {Murray}, {Nagai}, {Quintana}, \&
  {Voevodkin}}]{2009ApJ...692.1033V}
{Vikhlinin}, A., {Burenin}, R.~A., {Ebeling}, H., {et~al.} 2009, \apj, 692,
  1033

\bibitem[{{Wen} {et~al.}(2010){Wen}, {Han}, \& {Liu}}]{2010MNRAS.407..533W}
{Wen}, Z.~L., {Han}, J.~L., \& {Liu}, F.~S. 2010, \mnras, 407, 533

\bibitem[{{Wen} {et~al.}(2012){Wen}, {Han}, \& {Liu}}]{2012ApJS..199...34W}
{Wen}, Z.~L., {Han}, J.~L., \& {Liu}, F.~S. 2012, \apjs, 199, 34

\bibitem[{{Willmer} {et~al.}(2006){Willmer}, {Faber}, {Koo}, {Weiner},
  {Newman}, {Coil}, {Connolly}, {Conroy}, {Cooper}, {Davis}, {Finkbeiner},
  {Gerke}, {Guhathakurta}, {Harker}, {Kaiser}, {Kassin}, {Konidaris}, {Lin},
  {Luppino}, {Madgwick}, {Noeske}, {Phillips}, \& {Yan}}]{2006ApJ...647..853W}
{Willmer}, C.~N.~A., {Faber}, S.~M., {Koo}, D.~C., {et~al.} 2006, \apj, 647,
  853

\bibitem[{{Wolf} {et~al.}(2003){Wolf}, {Meisenheimer}, {Rix}, {Borch}, {Dye},
  \& {Kleinheinrich}}]{2003A&A...401...73W}
{Wolf}, C., {Meisenheimer}, K., {Rix}, H.-W., {et~al.} 2003, \aap, 401, 73

\bibitem[{{Wright} \& {Brainerd}(2000)}]{2000ApJ...534...34W}
{Wright}, C.~O. \& {Brainerd}, T.~G. 2000, \apj, 534, 34

\bibitem[{{York} {et~al.}(2000){York}, {Adelman}, {Anderson}, {Anderson},
  {Annis}, {Bahcall}, {Bakken}, {Barkhouser}, {Bastian}, {Berman}, {Boroski},
  {Bracker}, {Briegel}, {Briggs}, {Brinkmann}, {Brunner}, {Burles}, {Carey},
  {Carr}, {Castander}, {Chen}, {Colestock}, {Connolly}, {Crocker}, {Csabai},
  {Czarapata}, {Davis}, {Doi}, {Dombeck}, {Eisenstein}, {Ellman}, {Elms},
  {Evans}, {Fan}, {Federwitz}, {Fiscelli}, {Friedman}, {Frieman}, {Fukugita},
  {Gillespie}, {Gunn}, {Gurbani}, {de Haas}, {Haldeman}, {Harris}, {Hayes},
  {Heckman}, {Hennessy}, {Hindsley}, {Holm}, {Holmgren}, {Huang}, {Hull},
  {Husby}, {Ichikawa}, {Ichikawa}, {Ivezi{\'c}}, {Kent}, {Kim}, {Kinney},
  {Klaene}, {Kleinman}, {Kleinman}, {Knapp}, {Korienek}, {Kron}, {Kunszt},
  {Lamb}, {Lee}, {Leger}, {Limmongkol}, {Lindenmeyer}, {Long}, {Loomis},
  {Loveday}, {Lucinio}, {Lupton}, {MacKinnon}, {Mannery}, {Mantsch}, {Margon},
  {McGehee}, {McKay}, {Meiksin}, {Merelli}, {Monet}, {Munn}, {Narayanan},
  {Nash}, {Neilsen}, {Neswold}, {Newberg}, {Nichol}, {Nicinski}, {Nonino},
  {Okada}, {Okamura}, {Ostriker}, {Owen}, {Pauls}, {Peoples}, {Peterson},
  {Petravick}, {Pier}, {Pope}, {Pordes}, {Prosapio}, {Rechenmacher}, {Quinn},
  {Richards}, {Richmond}, {Rivetta}, {Rockosi}, {Ruthmansdorfer}, {Sandford},
  {Schlegel}, {Schneider}, {Sekiguchi}, {Sergey}, {Shimasaku}, {Siegmund},
  {Smee}, {Smith}, {Snedden}, {Stone}, {Stoughton}, {Strauss}, {Stubbs},
  {SubbaRao}, {Szalay}, {Szapudi}, {Szokoly}, {Thakar}, {Tremonti}, {Tucker},
  {Uomoto}, {Vanden Berk}, {Vogeley}, {Waddell}, {Wang}, {Watanabe},
  {Weinberg}, {Yanny}, {Yasuda}, \& {SDSS Collaboration}}]{2000AJ....120.1579Y}
{York}, D.~G., {Adelman}, J., {Anderson}, Jr., J.~E., {et~al.} 2000, \aj, 120,
  1579

\end{thebibliography}

\end{document}